\documentclass[usenatbib,usegraphicx]{mn2e}
\usepackage{amssymb}

\def\Lya{Ly$\alpha\ $}

\def\HI{\hbox{H~$\scriptstyle\rm I\ $}}

\def\msun{{M_\odot}}

\def\ltsima{$\; \buildrel < \over \sim \;$}
\def\lsim{\lower.5ex\hbox{\ltsima}}
\def\gtsima{$\; \buildrel > \over \sim \;$}
\def\gsim{\lower.5ex\hbox{\gtsima}}

\def\spose#1{\hbox to 0pt{#1\hss}}
\def\lta{\mathrel{\spose{\lower 3pt\hbox{$\mathchar"218$}}
     \raise 2.0pt\hbox{$\mathchar"13C$}}}
\def\gta{\mathrel{\spose{\lower 3pt\hbox{$\mathchar"218$}}
     \raise 2.0pt\hbox{$\mathchar"13E$}}}

\journal{Preprint-00}

\title{Galactic wind X-ray heating of the intergalactic medium during the Epoch of Reionization}

\author[ A. Meiksin, S. Khochfar, J.-P. Paardekooper, C. Dalla Vecchia, S. Kohn]{Avery Meiksin$^{1}$\thanks{E-mail:\ A.Meiksin@ed.ac.uk (AM)}, Sadegh Khochfar$^{1}$, Jan-Pieter Paardekooper$^{2}$, \newauthor Claudio Dalla Vecchia$^{3,4}$, Saul Kohn$^{5}$\\ 
$^{1}$SUPA\thanks{Scottish Universities Physics Alliance}, Institute for Astronomy, University of Edinburgh, Blackford Hill, Edinburgh\ EH9\ 3HJ, UK\\
$^{2}$Universit\"at Heidelberg, Zentrum f\"ur Astronomie, Institut f\"ur Theoretische Astrophysik, Albert-Ueberle-Str. 2, 69120 Heidelberg, Germany\\
$^{3}$Instituto de Astrofs\'{i}ca de Canarias, C/ V\'{i}a L\'{a}ctea s/n,38205 La Laguna, Tenerife, Spain\\
$^{4}$Departamento de Astrofs\'{i}ca, Universidad de La Laguna, Av. del Astrof\'{i}sico Franciso S\'{a}nchez s/n, 38206 La Laguna, Tenerife, Spain\\
$^{5}$Center for Particle Cosmology, Department of Physics and Astronomy, University of Pennsylvania, Philadelphia, PA\ 19104, USA}

\begin{document}

\maketitle

\begin{abstract}
  The diffuse soft X-ray emissivity from galactic winds is computed
  during the Epoch of Reionization (EoR). We consider two analytic
  models, a pressure-driven wind and a superbubble model, and a 3D
  cosmological simulation including gas dynamics from the First
  Billion Years (FiBY) project. The analytic models are normalized to
  match the diffuse X-ray emissivity of star-forming galaxies in the
  nearby Universe. The cosmological simulation uses physically
  motivated star formation and wind prescriptions, and includes radiative
  transfer corrections. The models and the simulation all are found to
  produce sufficient heating of the Intergalactic Medium to be
  detectable by current and planned radio facilities through 21~cm
  measurements during the EoR. While the analytic models predict a
  21~cm emission signal relative to the Cosmic Microwave Background
  sets in by $z_{\rm trans}\simeq8-10$, the predicted signal in the
  FiBY simulation remains in absorption until reionization
  completes. The 21~cm absorption differential brightness temperature
  reaches a minimum of $\Delta T\simeq-130$ to $-200$~mK, depending on
  model. Allowing for additional heat from high mass X-ray binaries
  pushes the transition to emission to $z_{\rm trans}\simeq10-12$,
  with shallower absorption signatures having a minimum of $\Delta
  T\simeq-110$ to $-140$~mK. The 21~cm signal may be a means of
  distinguishing between the wind models, with the superbubble model
  favouring earlier reheating. While an early transition to emission
  may indicate X-ray binaries dominate the reheating, a transition to
  emission as early as $z_{\rm trans}>12$ would suggest the presence
  of additional heat sources.
\end{abstract}


\section{Introduction}
\label{sec:intro}

Following the Recombination Era at $z\simeq1100$, the baryons produced
in the Big Bang cooled primarily by adiabatic expansion until the
first objects began to form. An important consequence is that the
21~cm signal from collapsed objects against the Cosmic Microwave
Background (CMB) will produce a strong absorption signal, varying
inversely with the gas temperature \citep{1979MNRAS.188..791H}. More
rarefied gas will produce an absorption signature against the CMB only
over a restricted redshift window when collisional coupling of the
spin temperature to the gas temperature dominates over radiative
coupling to the CMB \citep{1990MNRAS.247..510S}. At later times, the
more rarefied gas becomes invisible against the CMB until the first
luminous objects build up a \Lya\ radiation field sufficiently intense
to recouple the spin temperature of the neutral hydrogen to the gas
kinetic temperature through the Wouthuysen-Field (W-F) mechanism
\citep{1952AJ.....57R..31W, 1959ApJ...129..551F}. This will occur
during the Epoch of Reionization (EoR)
\citep{1997ApJ...475..429M}. This coincidence in times has motivated a
surge of interest in low-frequency radio arrays to search for the EoR
using the redshifted 21~cm line. These include extensions of previous
facilities like the Giant Metrewave Radio Telescope
\citep[GMRT;][]{2011MNRAS.413.1174P} and the design of a new
generation of arrays including the LOw Frequency ARray
\citep[LOFAR;][]{2013A&A...556A...2V}, the Murchison Widefield Array
\citep[MWA;][]{2013PASA...30....7T} and the Precision Array Probing
the Epoch of Reionization \citep[PAPER;][]{2010AJ....139.1468P}. A
second generation of instruments able to probe the small angular and
frequency scales on which the 21~cm signal may be particularly strong
compared with the background is currently under development, including
the Square Kilometre Array \citep[SKA;][]{2015aska.confE...1K} and the
Hydrogen Epoch of Reionization Array
\citep[HERA;][]{2017PASP..129d5001D}.

Provided the gas kinetic temperature remains below the CMB
temperature, both the overdense and diffuse gas components will
produce absorption signals once the W-F mechanism is active. If young
stars in galaxies are the dominant source of reionizing radiation, the
associated soft X-rays from supernovae remnants, X-ray binaries, and
possibly shock-heated coronal gas in the gravitationally collapsed
environment of the galaxies will sufficiently warm the diffuse
component to produce 21~cm radiation in emission rather than
absorption during the reionization epoch for an X-ray to optical
luminosity ratio $L_x/L_{\rm opt}>0.01$ \citep{1997ApJ...475..429M,
  2004ApJ...602....1C}, a value typical in the nearby Universe.

A more refined estimate may be made using the observed correlation
between the X-ray luminosity and star formation rate in nearby
galaxies \citep{2003MNRAS.340..210G, 2003MNRAS.339..793G,
  2009MNRAS.394.1741O, 2010ApJ...724..559L, 2012MNRAS.419.2095M,
  2012MNRAS.426.1870M, 2014MNRAS.437.1698M}. Because of uncertainties
in the spectrum of the escaping X-rays and in the star formation rate
histories of the first forming galaxies, a wide range of heating
scenarios have been put forward, with the transition from absorption to emission relative to the CMB occuring over a redshift range covering $12<z<20$
\citep{2006MNRAS.371..867F, 2006MNRAS.367..259H, 2011AnA...528A.149M,
  2012ApJ...760....3M, 2013MNRAS.431..621M, 2017MNRAS.469.1166D}.

It is unknown how accurate an extrapolation from the nearby Universe
to the epoch of the first galaxies is. The principal sources of the
X-rays are mass-transfer binaries and galactic winds. At early times,
the time to evolve an accreting low mass X-ray binary system is likely
too long for many to contribute to intergalactic heating. This is in
contrast to high mass X-ray binary systems (HMXBs), for which the
evolutionary timescale is much shorter. The difficulty in separating
emission from low mass and high mass X-ray binaries in nearby galaxies
introduces potentially large uncertainty into the high mass X-ray
binary contribution \citep{2012MNRAS.419.2095M,
  2010ApJ...724..559L}. Given the uncertainty in the amount of
internal absorption within galaxies, the diffuse soft X-ray emission
($0.5-2$~keV) from winds may range from about one third of the X-ray
binary contribution to dominating the emission
\citep{2012MNRAS.426.1870M}. Stellar population modelling of galaxies
and the evolution of their binary populations during the EoR, however,
suggests HMXBs alone may not heat the still neutral IGM to
temperatures above the CMB until $z\lta10$
\citep{2017ApJ...840...39M}, as in late reionization scenarios
\citep{2014Natur.506..197F}. Phenomenological modelling of the
expected X-ray emission similarly suggests late heating above the CMB,
not commencing until as late as $z\lta12$ and possibly with the IGM
temperature not exceeding the CMB temperature until after reionization
has substantially completed \citep{2017MNRAS.464.1365M}.

These estimates are all subject to uncertainties in the structure of
young galaxies during the EoR. If the environment of HMXBs in young
galaxies is much denser or dustier than in present day galaxies, local
absorption could substantially suppress the fraction of soft X-rays
that escape \citep[but see][]{2017MNRAS.469.1166D}. On the other hand,
the supernovae that drive galactic winds should be ubiquitous in any
star-forming region, leading to the production of a diffuse X-ray
component. If the escape fraction of soft X-rays is as low as in
present day galaxies, then the soft X-rays from galactic winds are
unlikely to contribute much to the reheating of the IGM. The early
galaxies, however, will be much less massive than present day
galaxies, and are expected to have substantially larger UV escape
fractions \citep{2006MNRAS.371L...1I, 2012ApJ...746..125H}. Much large
diffuse X-ray escape fractions may then be expected as well, so that
the diffuse X-ray emission from winds may play an important role in
heating the IGM \citep{2014MNRAS.443..678P}. As the metallicities and
star-formation rates will be different from present-day massive
galaxies, the X-ray spectra measured in nearby galaxies may not be
representative of the spectra of this early population, as has been
assumed in the literature. For these reasons we instead model the
expected X-ray emission from the galaxies.

An important uncertainty in modelling the X-ray contribution from
winds, however, is the driving mechanisms of the galactic winds, which
are still not fully understood. While the collective action of
supernovae in star-forming regions will drive a wind
\citep{1985Natur.317...44C}, agreement with early X-ray measurements
requires the wind to be heavily mass loaded so as not to overheat the
gas \citep{1993PASJ...45..513T}. Agreement with more recent X-ray data
continues to support this conclusion, and further suggests the amount
of mass loading increases with decreasing star formation rate
\citep{2014ApJ...784...93Z, 2016MNRAS.461.2762M}, and so may be
particularly large for galaxies during the reionization epoch. An
alternative model allows for the collective winds from supernovae in
stellar associations to form a superbubble which expands into an
ambient medium \citep{1987ApJ...317..190M,
  1988ApJ...324..776M}. Unlike for the freely-expanding wind solution,
which tacitly suppresses thermal heat conduction, the internal heat of
a superbubble diffuses so that the interior temperature equilibrates
with the shock front. Mass loading results from thermal evaporation
off the cavity walls. Other wind-driving mechanisms have been
suggested, including cosmic ray streaming \citep{1975ApJ...196..107I,
  2012MNRAS.423.2374U} and the momentum from radiation pressure and
supernovae \citep{2005ApJ...618..569M}.

These models have been incorporated into cosmological simulations to
approximate the \lq\lq sub-grid\rq\rq physics of feedback, necessary
to regulate the star formation rate in the simulations. While most
invoke energy-driven (ie, thermal pressure-driven), winds by
supernovae \citep[eg][]{2015MNRAS.446..521S, 2013MNRAS.436.3031V},
others allow for momentum-driven winds \citep{2014MNRAS.445..581H},
winds formed from superbubbles \citep{2015MNRAS.453.3499K}, and winds
driven by cosmic-ray pressure \citep{2014ApJ...797L..18S,
  2016ApJ...824L..30P}.

In this paper, we explore the expected strength of the X-ray component
due to supernovae, its dependence on galactic environment and its
possible contribution to the heating of the IGM before the Epoch of
Reionization completes. We consider both pressure-driven winds and
superbubbles, using approximate models to estimate the metagalactic
X-ray emissivity from forming galaxies. We also use a simulation drawn
from the First Billion Years (FiBY) suite of simulations of the first
galaxies \citep{2015MNRAS.451.2544P}, to estimate their contribution
to the X-ray emissivity during the EoR.

This paper is organized as follows:\ In the next section we estimate
the X-ray emissivity and IGM heating from supernovae in star-forming
galaxies for both analytic models and from the FiBY simulation. In the following section we compute the implications for the temperature evolution of the IGM. We then discuss the results and conclude.

Unless stated otherwise, the cosmological parameters adopted are
$\Omega_m=0.315$ for the total mass parameter, $\Omega_v=0.685$ for
the vacuum energy density, $\Omega_bh^2=0.0222$ for the baryon mass
density, where, for a present-day Hubble constant $H_0$, $h=H_0/ 100\,{\rm km\,s^{-1}\,Mpc^{-1}}=0.6731$. A $\Lambda$CDM power spectrum is assumed, normalized to $\sigma_8=0.829$ with $n_s=0.9655$. These values are consistsent with the 2015 {\it Planck} constraints \citep{2016A&A...594A..13P}.

\section{Diffuse soft X-ray emission estimates}
\label{sec:estimates}

\subsection{X-ray heating by supernovae}
\label{subsec:sources}

Supernovae in star forming regions will produce X-rays through thermal
bremsstrahlung and radiative recombination losses in hot gas. We begin
by considering two simplified wind models for estimating the amount of
X-ray heating supernovae in young galaxies may generate. Both are
based on the collective effects of supernovae that drive an expanding
sphere of hot gas. If the external pressure is negligible, a
freely-expanding wind is formed. We use the steady-state solution of
\citet{1985Natur.317...44C} to describe the structure of the wind. We
also consider the superbubble model. We use the models of
\citet{1987ApJ...317..190M} and \citet{1988ApJ...324..776M} to
describe the interior structure of a superbubble. Eventually the
superbubble will emerge vertically from the interstellar medium of the
galaxy and drive a wind into the halo \citep{1989ApJ...337..141M}.

For the analytic models and the FiBY simulation, we compute the
continuum X-ray emission from free-free and free-bound radiation from
hydrogen and helium, and the X-ray emission lines from metal
ions. X-rays softer than 200~eV will be photoelectrically absorbed
within 200--300~kpc (proper) of the sources, corresponding to the
average spacing between the sources, for a comoving spatial density of
about $0.01\,{\rm Mpc}^{-3}$ \citep{2015ApJ...803...34B}, while
photons with energies exceeding $\sim1$~keV have a proper mean free
path to photoelectric absorption by neutral hydrogen and helium
corresponding to a redshift of $\Delta z\sim1$ for $z>8$, at which
redshifting and evolutionary effects become important:
\begin{equation}
\lambda_x^{\rm prop}\simeq(63\,{\rm
  Mpc})\left(\frac{1+z}{9}\right)^{-3}E_{\rm keV}^{3.2},
\label{eq:lamx}
\end{equation}
for X-rays with energies $E$ (keV). We consider X-ray energies in the
range $0.2-1.6$~keV, extending the upper limit slightly, noting that
the emission is dominated by lower energy X-rays. If the population of
sources dominating reionization has a much higher comoving space
density than $0.01\,{\rm Mpc}^{-3}$, the lower energy limit would be
reduced and could result in significantly more X-ray heating. We
compute the metal emission line contribution by interpolating on
tables constructed using CLOUDY \citep[13.03][]{2013RMxAA..49..137F}
with the CHIANTI rates for collisionally ionized gas
\citep{2012ApJ...744...99L}, assuming solar abundance ratios. It is
possible $\alpha$-enhanced abundances are more relevant for an early
population of supernovae \citep{2012ApJ...744...91B}, however this
introduces only a small uncertainty for low metallicity gas.

In the absence of a complete theory for the generation and propagation
of galactic winds, for the simplified models considered here we rely
on observations for guidance \citep{2009MNRAS.394.1741O,
  2012MNRAS.426.1870M, 2014MNRAS.437.1698M}. There is some ambiguity
in the soft X-ray emission because of uncertainty in internal
absorption corrections for the observed galaxies. Without any
correction, \citet{2012MNRAS.426.1870M} obtain for the energy in
diffuse radiation between $0.5-2$~keV an amount
$\sim2\times10^{46}$~erg per solar mass of stars formed, and a
bolometric value of $\sim3\times10^{46}$~erg per solar mass. For a
subset of the galaxies in the sample, the spectral fit is improved
allowing for internal absorption. Using the best fit X-ray
luminosities for these galaxies gives an intrinsic bolometric emission
energy of $\sim5\times10^{47}$~erg per solar mass of stars
formed. Thus the escaping radiation may be reduced by an order of
magnitude by internal extinction. Alternatively, the lower value may
represent the intrinsic luminosity of most galaxies, presuming
extinction corrections are negligible within them. For the main body
of this paper, we choose the higher value for the intrinsic emission
level. We also note that any internal extinction corrections for the
high redshift galaxies could be very different from those inferred for
the nearby galaxy sample, which itself is uncertain. Indeed, the FiBY
simulations generally find rather small extinction corrections for the
soft X-rays from the interstellar gas in young galaxies during the
reionization epoch, consistent with the expectation if the IGM was
reionized by galaxies \citep{2006MNRAS.371L...1I,
  2012ApJ...746..125H}. For the analytic models, we generally assume
negligible internal extinction over $0.2-1.6$~keV. The values
discussed in the main text therefore provide high estimates of the
heating rate, most favourable for early heating of the still neutral
intergalactic gas. In the Appendix, we show that if the lower rates
are adopted, either because the intrinsic emission from galaxies is
lower or because of a high level of internal extinction, diffuse soft
X-ray emission from galaxies will heat the still neutral gas to
temperatures above the CMB only at very late times. For the FiBY
simulation, we compute the escape fractions directly through radiative
transfer of the soft X-rays.

It is instructive first to estimate the expected amount of soft X-ray
radiation emitted by a single supernova remnant expanding into a
surrounding medium of hydrogen density $n_{\rm H, 0}$, with
helium-to-hydrogen number ratio ${\rm He}/{\rm H}\simeq0.08$ and
metallicity $\zeta_m=Z/Z_\odot$. The supernova will produce a hot
cavity of X-ray emitting gas. For a typical core-collapse ejecta
mechanical energy of $E=10^{51}E_{\rm 51}\,{\rm erg}$, the radius of
the cavity in the Sedov-Taylor blast-wave phase expands with time
$t=10^6t_6$~yr like
\begin{equation}
R_{\rm S-T}\simeq 79\,{\rm pc}\left(\frac{E_{51}t_6^2}{n_{\rm H,0}}\right)^{1/5}
\label{eq:ST_rad}
\end{equation}
with post-shock temperature

\begin{equation}
T_{\rm sh} \simeq (6.6\times10^9\,{\rm K})\frac{E_{51}}{R_{\rm S-T}^3n_{\rm H,0}}.
\label{eq:ST_Tsh}
\end{equation}

The shell of the blast wave will cool due to continuum radiation from
hydrogen and helium and line radiation from metals. Once most of the
mechanical energy has been radiated away, the supernova remnant enters
a pressure-driven snowplough phase. The characteristic cooling time is
$t_{\rm PDR}\simeq(1.33\times10^4\,{\rm yr})E_{51}^{3/14}n_{\rm
  H,0}^{-4/7}(\zeta_m+0.15)^{-5/14}$ \citep{1988ApJ...334..252C}. The
term $0.15$ added to $\zeta_m$ approximately accounts for hydrogen and
helium cooling.

\begin{figure}
\includegraphics[width=3.3in]{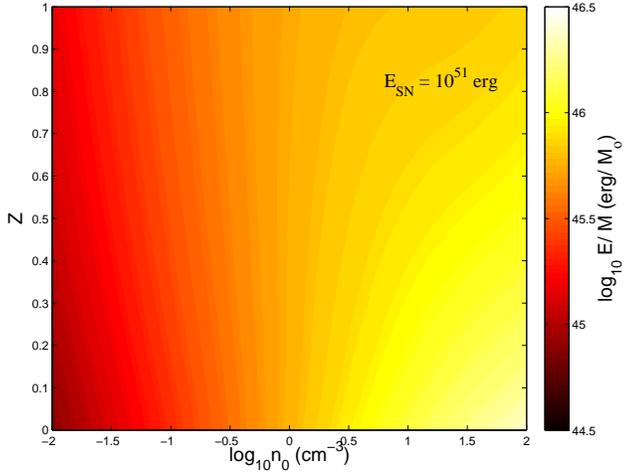}
\caption{X-ray emissivity per solar mass of stars formed produced by a single core-collapse supernova with energy $E=10^{51}\,{\rm erg}$ during the Sedov-Taylor expansion phase. Shown for X-ray energies integrated over $0.2-1.6$~keV.
}
\label{fig:EmissGridST}
\end{figure}

The X-ray emissivity in the range $0.2-1.6$~keV is shown in
Fig.~\ref{fig:EmissGridST} for a range of ambient hydrogen densities
and metallicities. The total energy radiated in the band is computed
using the Sedov-Taylor self-similar solution, integrated over time
until the remnant enters the snowplough phase. The emissivity has been
normalized to the mass in stars formed, assuming 1 supernova per 100
solar masses of stars formed, typical of a Salpeter stellar initial
mass function. The metallicity plays two opposing roles, decreasing
the cooling time while increasing the soft X-ray luminosity for
increasing metallicity. At low ambient gas densities, the remnant
expands until the temperature drops sufficiently for metal line
emission to contribute substantially to the soft X-rays emission, with
the total amount of energy emitted before the snowplough phase
increasing with increasing metallicity. At high gas densities, the
remnant cools rapidly, with the emission dominated by continuum
emission from hydrogen and helium. As the metallicity increases, the
cooling time shortens, resulting in less emission. A cross-over occurs
at an ambient gas density of $n_{\rm H,0}\simeq1\,{\rm cm^{-3}}$, for
which continuum emission dominates at low metallicity and metal line
emission at high, resulting in a total emissivity that is nearly
independent of metallicity. Allowing for a range in densities and
temperatures, the soft X-ray energy released per solar mass of stars
formed is between about $10^{44.5}-10^{46.5}\,{\rm erg\,
  M_\odot^{-1}}$, or a time-averaged soft X-ray luminosity of
$10^{37.0}-10^{39.0}\dot M_*\,{\rm erg\,s^{-1}}$ for the star
formation rate $\dot M_*$ in units of $M_\odot\,{\rm yr}^{-1}$. The
total energy in soft X-rays emitted per supernova exceeds
$10^{47.7}\,{\rm erg}$ for $n_{\rm H,0}>1\,{\rm cm^{-3}}$, or about
$0.0005$ of the blast energy. The higher values for the soft X-ray
energy radiated per solar mass of stars formed are comparable to the
measured values, assuming negligible extinction corrections
\citep{2009MNRAS.394.1741O, 2012MNRAS.426.1870M}.

Since early galaxies are also expected to form Pop~III stars, an
additional contribution may arise from Pair-Instability Supernovae
(PISNe), with progenitor masses of $140-260\,M_\odot$ and an average
mechanical ejecta energy of $3\times10^{52}$~ergs
\citep{2002ApJ...567..532H, 2013MNRAS.428.1857J}. The dependence of
the resulting soft X-ray emission from a remnant on ambient gas
density and metallicity is similar to that for a core-collapse
supernova, except the total energy released is increased by about a
factor of 30. Given that Pop~III star formation may not last long
before the gas is polluted by metals
\citep[eg][]{2011MNRAS.414.1145M}, and depending on the initial mass
function of the stars formed, the additional contribution from PISNe
may range from negligible \citep{2017MNRAS.467..802C}, to an amount
comparable to that of core-collapse supernovae.

An improved estimate may be made for the total X-ray emission from
supernova remnants by continuing the blast solutions into the
pressure-driven snowplough phase, however two complications arise:\
since stars form in associations further supernovae will explode
within the cavity, and in time separate cavities will collide and
begin to overlap. Both effects will delay cooling, resulting in an
increase in the total energy radiated per supernova. The overlapping
regions will moreover drive a galactic wind. It may then be expected
that galactic winds will have a higher productivity of X-ray emission
than isolated supernova remnants. We consider two simple wind models,
a pressure-driven wind in a steady state and an expanding superbubble.

\subsection{Pressure-driven wind}
\label{subsec:pwind}

\begin{figure}
\includegraphics[width=3.3in]{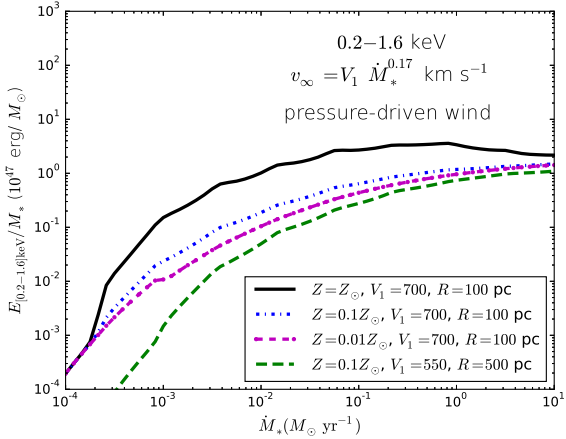}
\caption{X-ray emissivity per solar mass of stars formed within a region of radius $R$ by a pressure-driven wind with asymptotic wind velocity $v_\infty=V_1{\dot M}_*^{0.17}\,{\rm km\,s^{-1}}$. Shown for X-ray energies integrated over $0.2-1.6$~keV for the indicated metallicities. The emissivity decreases for decreasing metallicity or increasing size of the star forming region.
}
\label{fig:EmissEoRCCWind}
\end{figure}

A freely expanding pressure-driven wind produced by supernovae within
a star-forming region of radius $100R_{100}$~pc will approach an
asymptotic velocity $v_\infty$. In terms of the star-formation rate 
$\dot M_*$, in units of
$\msun\,{\rm yr}^{-1}$, once the interior of the wind achieves a
steady state the central temperature of the wind for a $\gamma=5/3$ gas is

\begin{equation}
  T_0 \simeq1.43\times10^7\,{\rm K}\,\frac{\chi}{\eta}
      \simeq14.2v_\infty^2,
\label{eq:T0}
\end{equation}

\noindent \citep{1985Natur.317...44C, 2016MNRAS.461.2762M}, where
$\bar m$ is the mean mass per particle, $\chi$ characterizes the
energy injection into the wind, and $\eta=\dot M/\dot M_*$ is the
mass-loading factor of a wind with a mass flow rate $\dot M$. For a
typical mechanical energy of $10^{51}\,{\rm erg}$ per supernova and 1
supernova per $100\,M_\odot$ of stars formed, energy is injected at
the rate $\dot E\simeq3.17\times10^{41}\,{\rm erg\,s^{-1}}\,\chi{\dot
  M_*}$. The mass-loading factor is related to the asymptotic wind
velocity by
\begin{equation}
\eta\simeq \chi\left(\frac{1000\,{\rm km\,s^{-1}}}{v_\infty}\right)^2.
\label{eq:eta}
\end{equation}

Consistency with the measured linear correlation of the diffuse soft
X-ray luminosity of star-forming galaxies with the star-formation rate
may be achieved in the pressure-driven wind model for an asymptotic
wind velocity that scales approximately as
$v_\infty\simeq(700-1000)\,\dot M_*^{1/6}$
\citep{2016MNRAS.461.2762M}, with the lower value allowing for
internal galaxy absorption and the higher value assuming negligible
internal absorption. For low star formation rates, the asymptotic wind
velocity corresponds to substantial mass loadings which may drive the
temperatures to values too low to produce large X-ray luminosities. If
the mass loading is too great, it may suppress the wind altogether
through radiative cooling. For a metallicity $\zeta_m$ relative to
solar, the mass loading factor is restricted to
\begin{equation}
\eta < 3.3\chi^{0.73}\left(\zeta_m\dot M_*R_{100}^{-1}\right)^{-0.27}
\label{eq:etamax}
\end{equation}
\citep{2016MNRAS.461.2762M}. Compton cooling was found to have little
effect at the redshifts considered.

Direct measurements of the linewidths of absorption lines in nearby
winds suggest instead a broader range of wind velocities for a given
star formation rate. The outflow velocities $v_{\rm out}$ measured by
\citet{2015ApJ...809..147H} suggest the steeper correlation $v_{\rm
  out}\simeq 400\,\dot M_*^{1/3}\,{\rm km\,s^{-1}}$. Because the
absorption features are dominated by positions in the flow for which
the ion measured has its highest column density, which depends on the
gas temperature, it is unclear how to relate $v_{\rm out}$ and
$v_\infty$. Nonetheless, if the steeper correlation is accepted, then,
as shown in the Appendix, the X-ray emissivity in the
$2.3<\log_{10}E_{\rm eV}<3.2$ band is reduced precipitously at low
star formation rates, and peaks in the range $0.02<\dot
M_*<1\,M_\odot\,{\rm yr}^{-1}$. Only the scaling $v_\infty\sim \dot
M_*^{1/6}$ is considered here.

The intrinsic emissivity in the $2.3<\log_{10}E_{\rm eV}<3.2$ band per
solar mass of stars formed is shown in
Fig.~\ref{fig:EmissEoRCCWind}. The emissivity increases with star
formation rate, which corresponds to an increasing gas density in the
pressure-driven wind model according to $n_{\rm H,0}\sim\dot
M_*^{1/2}$ \citep{2016MNRAS.461.2762M}. This agrees with the
expectation for a single supernova remnant in the blastwave expansion
phase, for which the emissivity also increases with gas
density. Unlike in the blastwave case, however, the emission is not
curtailed by cooling. Instead a steady-state outflow is established,
with an emissivity that increases with metallicity. Except at very
small star formation rates, and rates exceeding about
$10\,M_\odot\,{\rm yr}^{-1}$, the emissivity is dominated by metal
line emission. A star forming region with radius 100~pc is assumed for
most cases. For a larger radius of 500~pc, the emission is
substantially reduced even though the asymptotic wind velocity adopted
reproduces the $0.3-10$~keV emissivity at low redshifts for this size
region (see the Appendix).

The rise in the stellar mass-specific emissivity with star-formation
rate and the sensitivity to the metals demonstrates the importance of
modelling the winds. The effects found here, especially for the low
star-formation rates, are missed in nearby galaxy samples, and would
be unaccounted for in models of X-ray heating during the EoR based on
adopting a fiducial X-ray spectrum from observations of nearby
galaxies.

\subsection{Superbubble}
\label{subsec:sbub}

\begin{figure}
\includegraphics[width=3.3in]{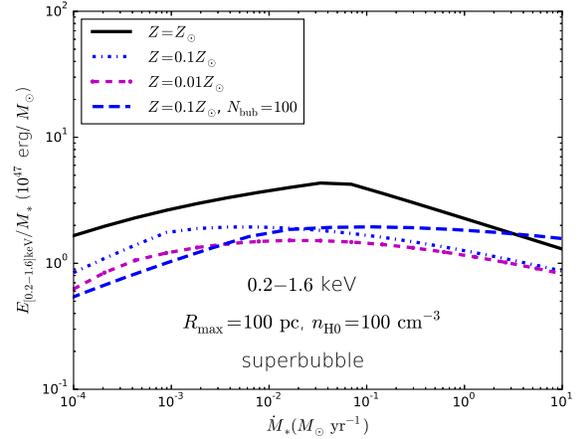}
\caption{X-ray emissivity per solar mass of stars formed from a superbubble reaching maximum radius $R_{\rm max}=100$~pc in an ambient medium of hydrogen density $n_{\rm H0}=100\,{\rm cm}^{-3}$. Also shown for a case in which the indicated star formation rate is shared equally between 100 superbubbles each of radius 100~pc. The X-ray emissivity is integrated over $0.2-1.6$~keV for the indicated metallicities. The emissivity decreases for decreasing metallicity but is otherwise nearly independent of the star formation rate. 
}
\label{fig:EmissEoRSB}
\end{figure}

The superbubble model describes the growth of a collective cavity of hot gas around a star-forming complex as the cavity expands into the surrounding interstellar medium. Mass loading is provided by thermal evaporation off the cavity wall. Although the superbubble may eventually develop into a wind as it grows in radius to a galactic disk scale height, we focus on the early stage here. The later stages are expected to be similar to the pressure-driven wind case.

A superbubble that expands to a maximum radius $R_{\rm max}=100R_{\rm B, 100}$~pc will have a central temperature
\begin{equation}
  T_c\simeq4.6\times10^7\,{\rm K}\left(\dot M_*/f_TR_{\rm B,100}\right)^{2/7}
  \label{eq:TcSB}
\end{equation}
\citep{2016MNRAS.461.2762M}, where suppression of thermal heat conduction by the factor $f_T$ has been allowed for. A supernova rate of 1 supernova per $100\,M_\odot$ of stars formed has been assumed, with a mechanical energy of $10^{51}$~erg injected into the surrounding gas per supernova. The temperature is independent of the ambient gas density, and only weakly dependent on the star formation rate, bubble size and thermal heat conduction suppression factor. While the bubble, or multiple bubbles that merge, may develop into a pressure-driven wind on expanding outside the disk of a galaxy and adiabatically cool, the component within the disk will maintain soft X-ray temperatures. Superbubbles are thus a robust source of soft X-rays. The free-free emission will be nearly flat in the range $0.2-1.6$~keV, corresponding to
\begin{eqnarray}
\frac{E_{[0.2-1.6]\,{\rm keV}}}{M_*}&\simeq&4\times10^{45}\,{\rm  
  erg}\,{M_\odot}^{-1}\,f_T^{5/7}\nonumber\\
&&\times n_{{\rm H},0}^{2/3}{\dot M_*}^{-8/21}R_{\rm B, 100}^{22/21},
\label{eq:ExMssswtc}
\end{eqnarray} 
\citep{2016MNRAS.461.2762M}
where $n_{\rm H0}$ is the ambient hydrogen density. Because the energy per star formed declines weakly with the star formation rate, the emissivity will be enhanced for a given net star formation rate if shared between $N_{\rm bub}$ superbubbles, all of the same maximum size, by the factor $N_{\rm bub}^{8/21}$. For hundreds of bubbles, this could boost the net energy released per star formed by factors of several.

For a sufficiently weak star formation rate, or sufficiently high
ambient gas density, the growth of the superbubble will be restricted
by radiative cooling. In this limit, the central temperature and
emissivity become
\begin{equation}
  T_c\simeq1.4\times10^7\,{\rm K}\left[n_{\rm H0}^2\dot M_*^2(\zeta_m+0.15)^3/f_T\right]^{1/11}
  \label{eq:TcSBclim}
\end{equation}
and
\begin{eqnarray}
\frac{E_{[0.2-1.6]\,{\rm keV}}}{M_*}&\simeq&2\times10^{47}\,{\rm  
  erg}\,{M_\odot}^{-1}\,(\zeta_m+0.15)^{-1},
\label{eq:ExMssswtcclim}
\end{eqnarray} 
\citep{2016MNRAS.461.2762M}, independent of the star formation rate, the ambient hydrogen density, the thermal heat conduction suppression factor, and nearly independent of the metallicity. Sharing the global star formation rate among multiple bubbles in this limit has no effect on the net emissivity, however the limit will be met more readily because of the reduced star formation rate per superbubble. A sufficiently weak star formation rate will result in a temperature cool enough to reduce the emissivity.

A range of superbubble model parameters matches the X-ray data in the
nearby Universe, as discussed in the Appendix. Typical parameters are
$R_{\rm max}=100$~pc and $n_{\rm H0}=100\,{\rm cm}^{-3}$. The soft
X-ray emission in the $0.2-1.6$~keV band is shown in
Fig.~\ref{fig:EmissEoRSB}. Comparable emissivities are obtained for
other combinations of the wind parameters when constrained to
reproduce the measured emissivity in the $0.3-10$~keV band in the
local Universe (see the Appendix).

As the metallicity is decreased, the transition to the cooling limit moves to smaller star formation rates. Generally the emissivity decreases at very low star formation rates because of the reduction in the temperature to energies comparable to or lower than the band energy. For the two cases shown with $Z=0.1Z_\odot$, allowing the star formation rate to be shared by $N_{\rm bub}=100$ bubbles boosts the net emissivity for $\dot M_*>0.01\,M_\odot\,{\rm yr}^{-1}$, but the cooling restriction reduces the rate at smaller star formation rates, until the low temperatures produce a decline in the band emissivity with declining star formation rate. 

\subsection{FiBY simulation X-ray emissivity}
\label{subsec:FiBY}

The FiBY simulations uses an updated version of the smoothed particle
hydrodynamics (SPH) code \texttt{GADGET} \citep[last described
  in][]{2005MNRAS.364.1105S}, including the modifications described in
\citet{2010MNRAS.402.1536S}. Details are provided in
\citet{2013MNRAS.428.1857J} and \citet{2013MNRAS.429L..94P,
  2015MNRAS.451.2544P}. Both Pop~II and Pop~III star formation is
incorporated along with energy-driven feedback from supernovae
following \citet{2012MNRAS.426..140D}, which reproduces the
Kennicutt-Schmidt law for star formation. This paper uses results from
the ${\rm FiBY\_S}$ simulation, run in a 4 Mpc (comoving) box with gas
particle mass $1254\,M_\odot$. The cosmological parameters used are
$\Omega_m=0.265$, $\Omega_v=0.735$, $\Omega_bh^2=0.0226$ with
$h=0.71$, and $\sigma_8=0.81$.

The X-ray emissivity is computed from the gas within the galaxies,
using the local metallicity, which ranges from highly subsolar to
solar (Dalla Vecchia, in preparation). The mean X-ray escape fraction
from a galaxy was computed following \citet{2015MNRAS.451.2544P},
using the \texttt{SIMPLEX} radiative transfer code
\citep{2010A&A...515A..79P}. The soft X-ray band was covered using 8
frequency bins. Metal absorption has not been included. For
metallicities above $0.1Z_\odot$, metal absorption will be small
compared with hydrogen and helium absorption, but dominates for solar
metallicity \citep{1983ApJ...270..119M}. If the soft X-rays escape
primarily through low-attenuation tunnels within the interstellar gas
of the galaxy, as does much of the photoionizing UV radiation
\citep{2015MNRAS.451.2544P}, even for solar metallicity the metals
will make little difference. In future, however, the effect of metals
on the escape fraction should be examined.

Only haloes more massive than about $10^6\,M_\odot$ were
resolved. Results are shown only for galaxies sufficiently gaseous to
perform radiative transfer computations, necessary for estimating the
escape fraction. This corresponds to a minimum gas mass per system of
about $10^5\,M_\odot$. For halo masses below $10^9\,M_\odot$, about
10--50 percent of virialized haloes meet this criterion, with the
fraction increasing towards higher mass haloes. For the redshifts of
interest, typically a few hundred to several hundred haloes were
included in the analysis at a given redshift. Further details on the
selection of the haloes are provided in \citet{2015MNRAS.451.2544P}.

\begin{figure*}
\scalebox{0.45}{\includegraphics{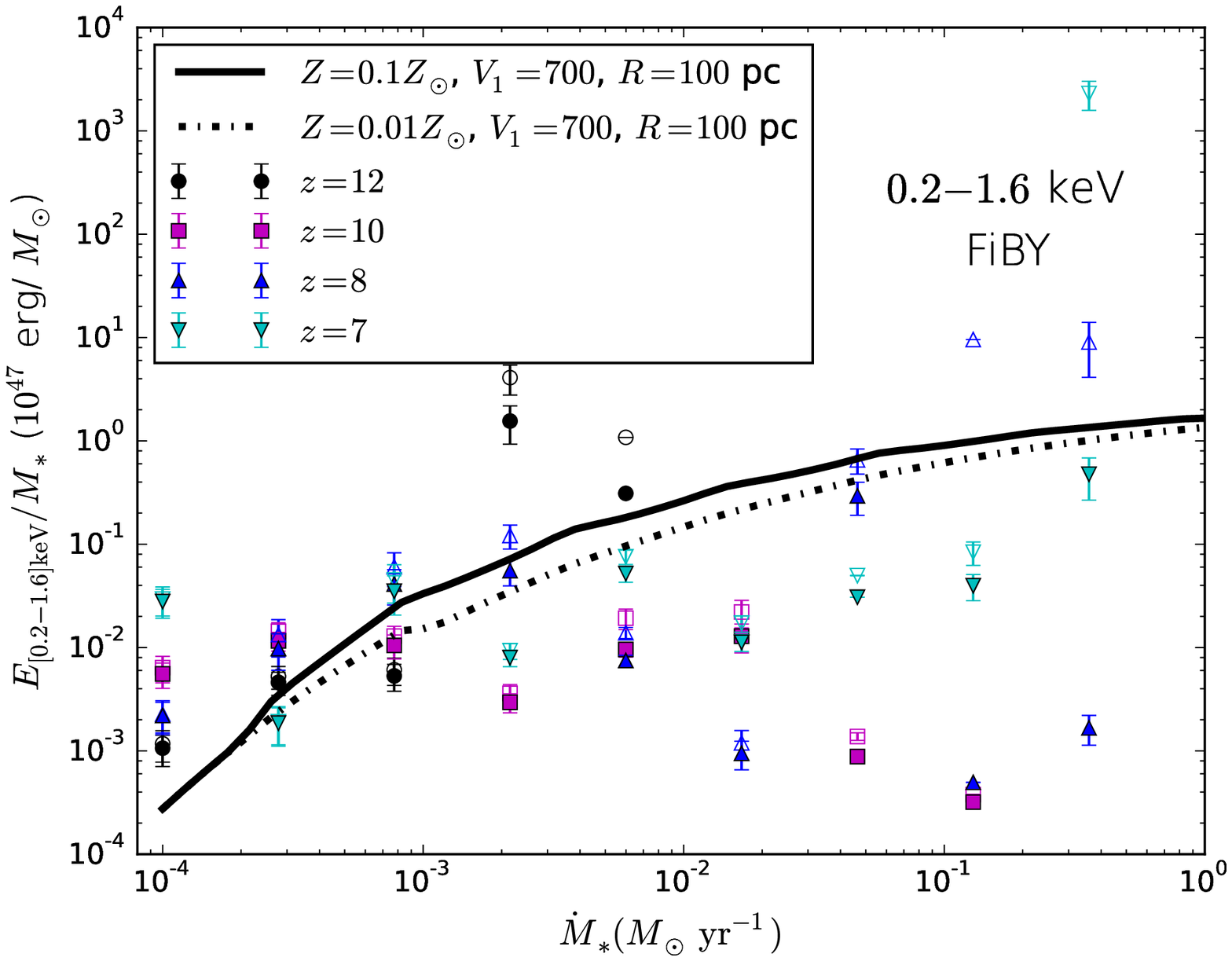}\includegraphics{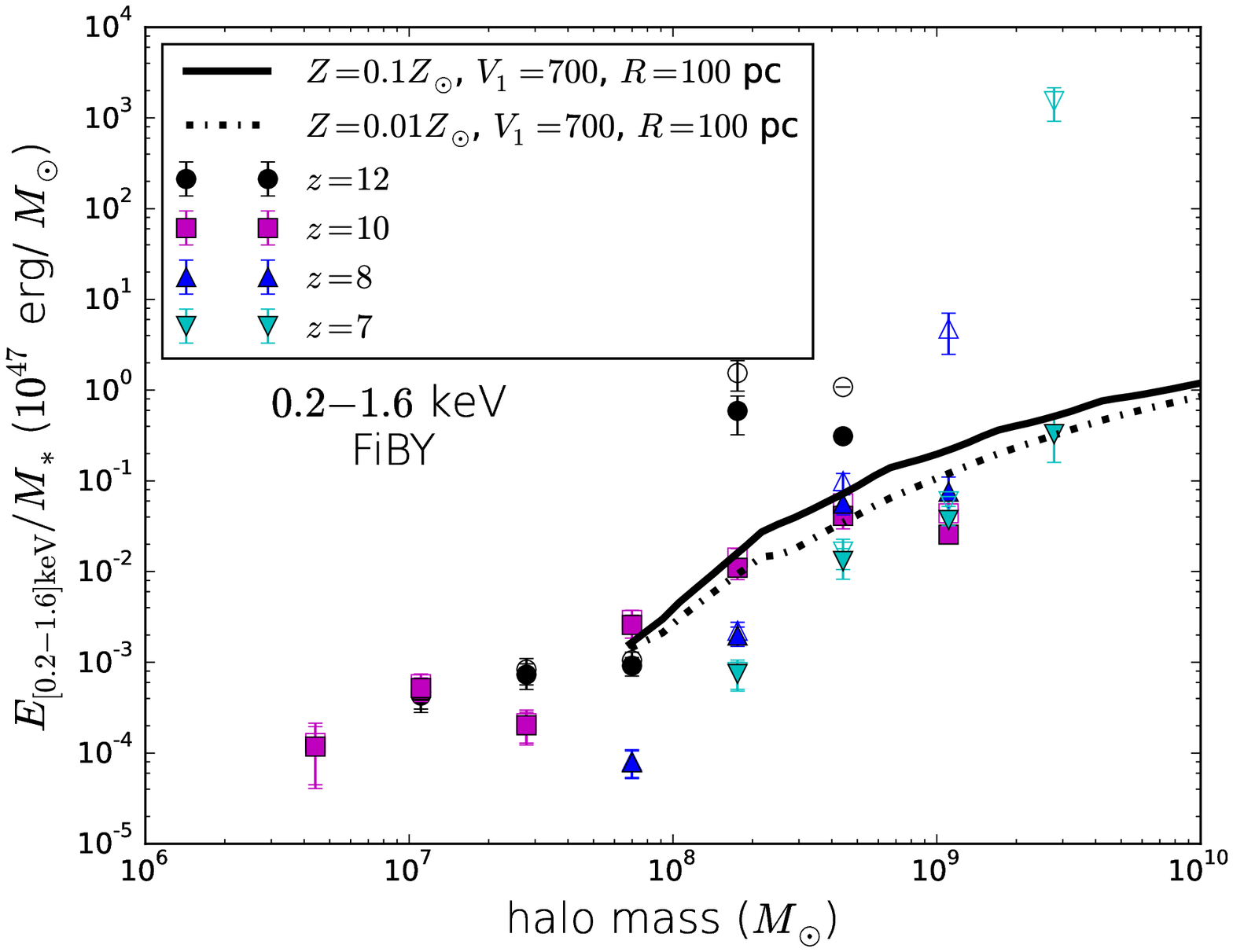}}
\caption{X-ray emissivity per solar mass of stars formed from the FiBY
  simulation, both without (open symbols) and including (solid
  symbols) the attenuation of soft X-rays by interestellar gas within
  the galaxies. Also shown is the case for a pressure-driven wind with
  bubble radius $R=100$~pc with asymptotic wind velocity
  $v_\infty=700\,{\rm km\,s^{-1}}\, \dot M_*^{0.17}$ and metallicities
  $0.01Z_\odot$ and $0.1Z_\odot$. The X-ray emissivity is integrated
  over $0.2-1.6$~keV. (Left panel) The emissivity is shown as a
  function of the star formation rate. While there is substantial
  scatter, the FiBY simulation emissivity declines for decreasing star
  formation rate following a trend similar to the pressure-driven wind
  model, especially for the lower redshifts shown. (Right panel) The
  emissivity from star-forming galaxies is shown as a function of halo
  mass. For the prediction of the pressure-driven wind model, star
  formation rates are converted to halo masses using the formalism of
  \citet{2015ApJ...813...21M}. The FiBY results follow the general
  trend predicted for a pressure-driven wind.
}
\label{fig:EmissEoRFiBY}
\end{figure*}

The emissivity per solar mass of stars formed in the X-ray band
$0.2-1.6$~keV is shown in Fig.~\ref{fig:EmissEoRFiBY}, both as a
function of star formation rate (left panel) and halo mass (right
panel), for star-forming haloes. The emission without attenuation of
soft X-rays by gas internal to the galaxies is shown by the open
symbols, while solid symbols include attenuation. The escape fraction
of soft X-rays ranges from less than one percent to near unity. Lower
escape fractions are generally found for galaxies with the larger star
formation rates, above $10^{-3}\,M_\odot\,{\rm yr}^{-1}$, or for the
more massive haloes, above $10^8\,M_\odot$. The attenuation, however, is highly
stochastic.

As a comparison, also shown is the prediction of a $Z=0.1Z_\odot$
pressure-driven wind model with asymptotic wind velocity scaling with
star formation rate set to match measured X-ray data in the
$0.3-10$~keV band \citep{2016MNRAS.461.2762M}. Order of magnitude
agreement with the FiBY results is found for $\dot
M_*<0.01\,M_\odot\,{\rm yr^{-1}}$, as shown in the left panel,
particularly at late times $z\le8$. In the right panel, star formation
rates have been converted to halo masses using the formalism of
\citet{2015ApJ...813...21M} from provided tables (C. Mason, personal
communication.) The tables are restricted to
$M_h>5\times10^7\,M_\odot$. The general trend with halo mass agrees
well with the FiBY results over $10^8-10^9\,M_\odot$.

\begin{figure}
\includegraphics[width=3.3in]{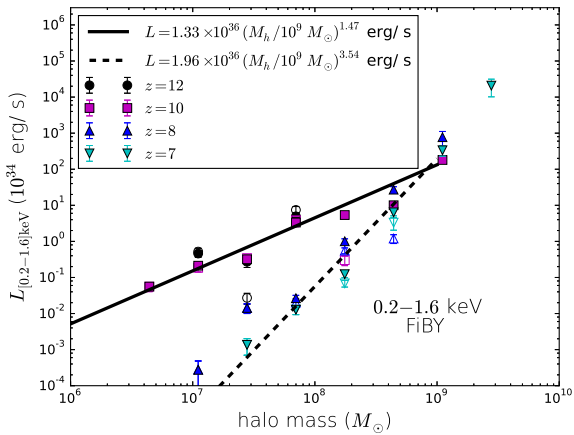}
\caption{X-ray luminosity as a function of halo mass from the FiBY
  simulation. The X-ray emissivity is integrated over
  $0.2-1.6$~keV. The filled symbols show the mean luminosity from all
  gaseous haloes while the open symbols show the mean luminosity from
  non-star forming gaseous haloes. Power-law fits are shown to the
  luminosities of gaseous haloes before (solid line) and after (dashed
  line) reionization at $z=9$.
}
\label{fig:L0EoRFiBY}
\end{figure}

The mean soft X-ray luminosities in the $0.2-1.6$~keV band for all
gaseous haloes, including corrections for attenuation, are shown in
Fig.~\ref{fig:L0EoRFiBY} as a function of halo mass. Both haloes
actively forming stars (solid symbols) and those not having formed
stars in the previous $5\times10^6$~yrs (open symbols) are shown. Two
general trends of increasing luminosity with halo mass are found, with
the trend steepening strongly after cosmic reionization at $z=9$, when
star formation is quenched in the lower mass haloes. Power-law fits
for the star-forming haloes with masses $M_h<10^9\,M_\odot$ are
$L_{[0.2-1.6]{\rm keV}}=1.33\times10^{36}\,{\rm
  erg\,s^{-1}}(M_h/10^9\,M_\odot)^{1.47}$
($z=10$) and
$L_{[0.2-1.6]{\rm keV}}=1.96\times10^{36}\,{\rm
  erg\,s^{-1}}(M_h/10^9\,M_\odot)^{3.54}$
($z=7$). The curves meet for halo masses $M_h\lta10^9\,M_\odot$. At
the low halo mass end, the mean X-ray luminosity of star-forming
haloes differs little from their non-star forming counterparts. The
residual X-ray luminosity may be the remnant of previous episodes of
star formation, but emission will also be produced by gas accreted by
the halo after shocking and becoming collisionally ionized.

\section{Cosmological soft X-ray heating rate}
\label{sec:cosxray}

The X-ray emission for the analytic models from galaxies in a
cosmological context may be estimated from the halo mass
function. Since the X-ray emission from the models depends on the star
formation rates, it is necessary to relate the star formation rate to
halo mass. The modelling of the galaxy luminosity function and its
evolution suggests such a relation is at best approximate
\citep{2013ApJ...770...57B}. We use the model of
\citet{2015ApJ...813...21M}, as it extends to high redshifts into the
expected reionization epoch. The model is non-unique, but provides a
concrete formalism that reproduces some average observed trends. We
allow for very low luminosity haloes, corresponding to the higher of
the two star formation rate extrapolations towards high redshifts
($z>10$) in \citet{2015ApJ...813...21M}. Allowing for a range in
uncertainties, the epoch of reionization was found to lie at
$z\simeq7.84^{+0.65}_{-0.98}$. As shown in the Appendix, the star
formation rate extrapolation agrees well with the rates found in the
FiBY simulation, although it is a source of uncertainty. It is
conservative in the sense that the lower extrapolation would produce
even less heating.

\subsection{Heating by pressure-driven wind}
\label{subsec:heatpwind}

\begin{figure*}
\scalebox{0.45}{\includegraphics{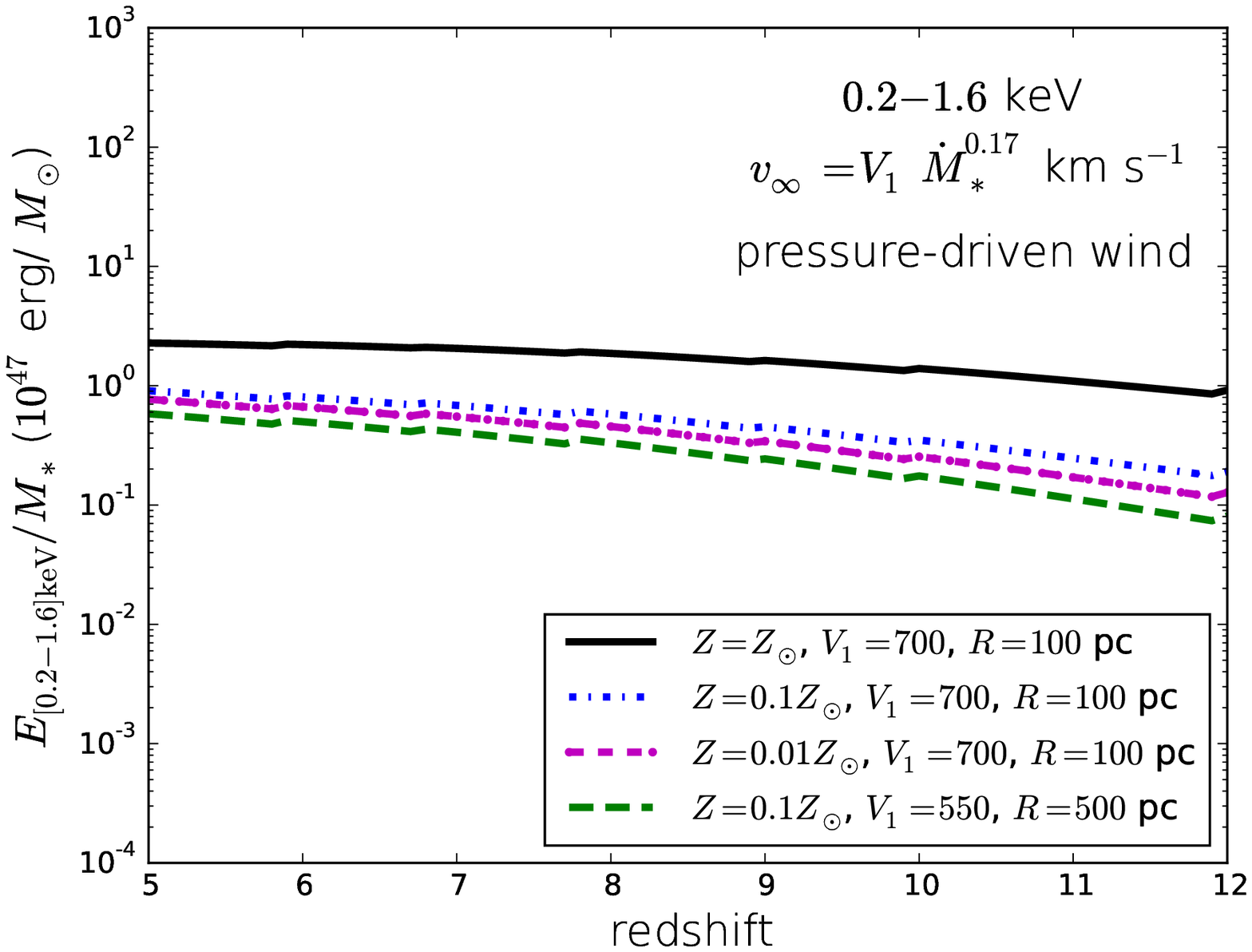}\includegraphics{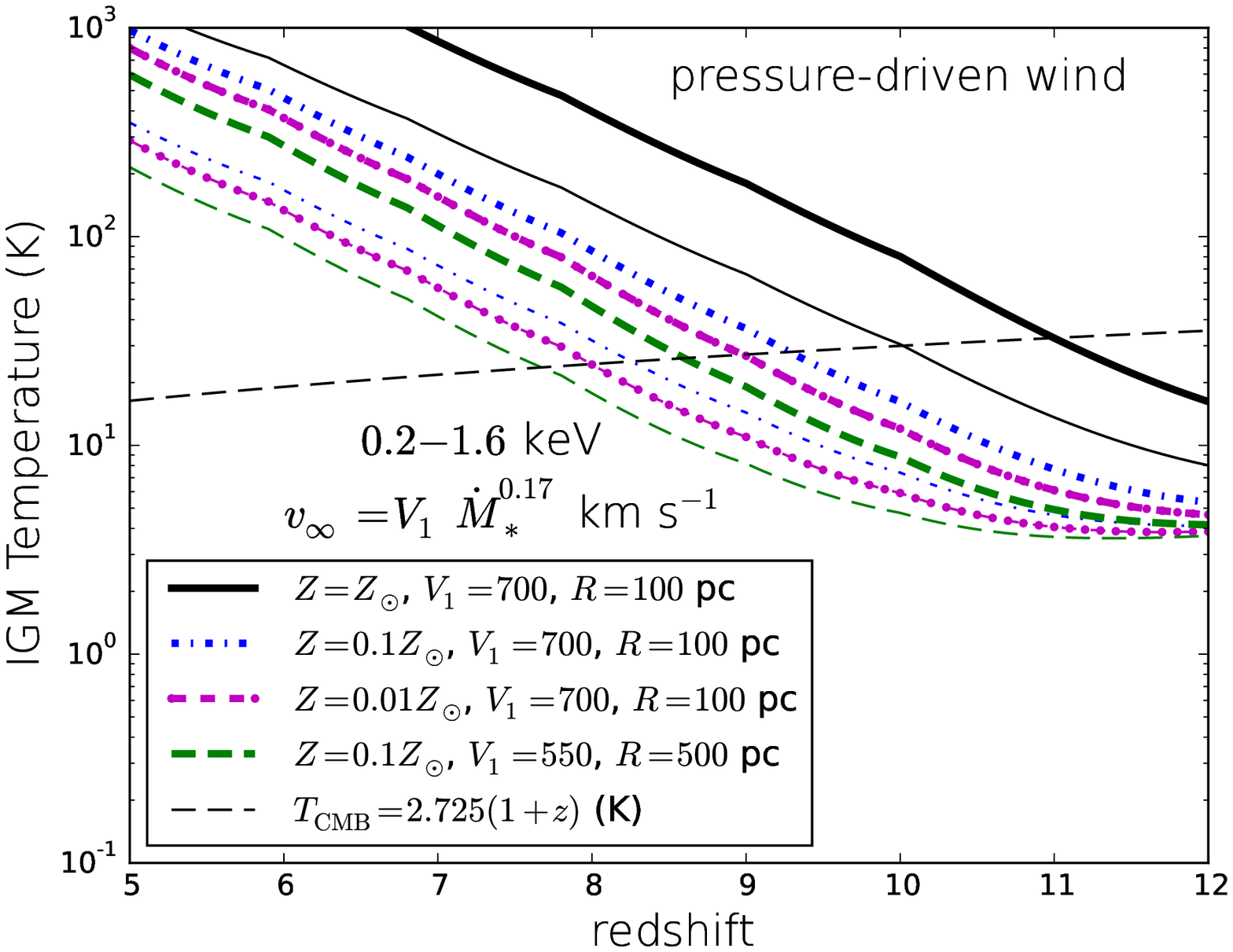}}
\caption{(Left panel) The halo-averaged soft X-ray emissivity as a function of redshift for a pressure-driven wind with asymptotic wind velocity correlated with the star formation rate according to $v_\infty=V_1\dot M_*^{0.17}\,{\rm km\, s^{-1}}$, for a variety of gas metallicities. (Right panel) The evolution in the temperature of the still neutral IGM using the X-ray emissivities in the left panel. The heavy lines assume all the X-ray energy is deposited in the IGM. The light lines allow for losses to secondary ionizations, with only 36 percent of the X-ray energy deposited as heat (see text). The dashed line rising towards higher redshifts is the CMB temperature.}
\label{fig:TEvolCCWind}
\end{figure*}

The net emissivity in the $0.2-1.6$~keV band summed over the haloes is
shown in the left panel of Fig.~\ref{fig:TEvolCCWind} for
pressure-driven winds. The asymptotic wind velocity correlation with
the star formation rate is as in Fig.~\ref{fig:EmissEoRCCWind}, and
corresponds to reproducing the measured correlation between X-ray
luminosity and star formation rate at low redshift
\citep{2016MNRAS.461.2762M}. For gas with solar metallicity, the rise
in the emissivity from $z=12$ to 5 is moderate, with most arising from
emission lines. At lower metallicities, the emissivity increases by
nearly an order of magnitude over this redshift interval. Most of the
emission is from free-free radiation, and is only weakly sensitive to
metallicity or size of the star-forming region.

The evolution of the temperature of the still neutral component of the IGM is shown in the right panel of Fig.~\ref{fig:TEvolCCWind}. The heavy lines assume all the soft X-ray energy is absorbed by the gas. Losses to secondary ionizations, however, will be substantial. \citet{2017ApJ...840...39M} estimate that at most 36 percent of the energy will be absorbed as heat within the neutral component as reionization proceeds. While winds with solar metallicity gas will warm the neutral component above the CMB temperature by $z\simeq11$ when all the soft X-ray energy is absorbed as heat, allowing for the reduced heating efficiency decreases this redshift to $z\simeq10$. For more plausible wind metallicities of $Z\sim0.1Z_\odot$, the reduced emissivity results in gas temperatures below the CMB temperature until $z\simeq8$.

\subsection{Heating by superbubbles}
\label{subsec:heatsbub}

\begin{figure*}
\scalebox{0.45}{\includegraphics{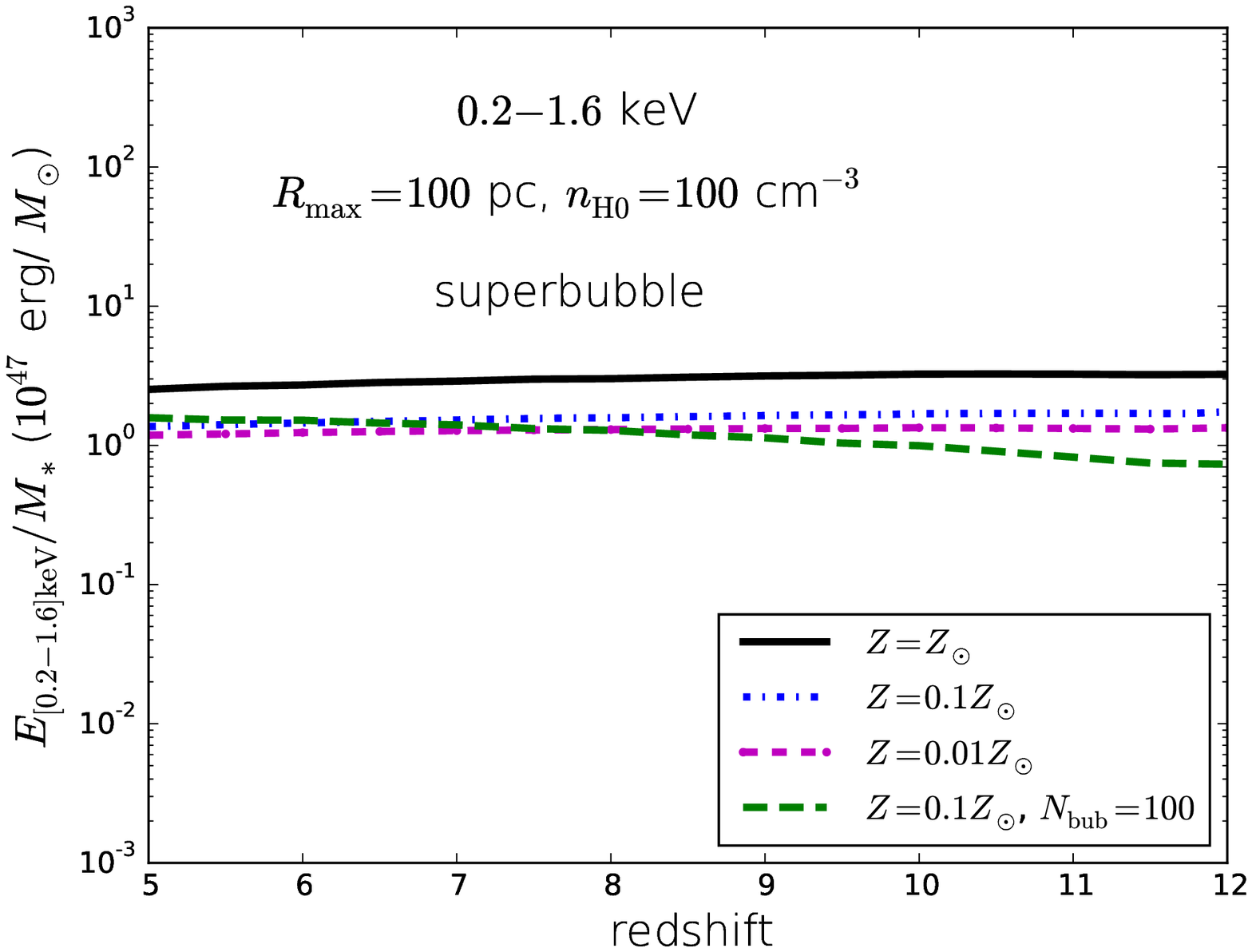}\includegraphics{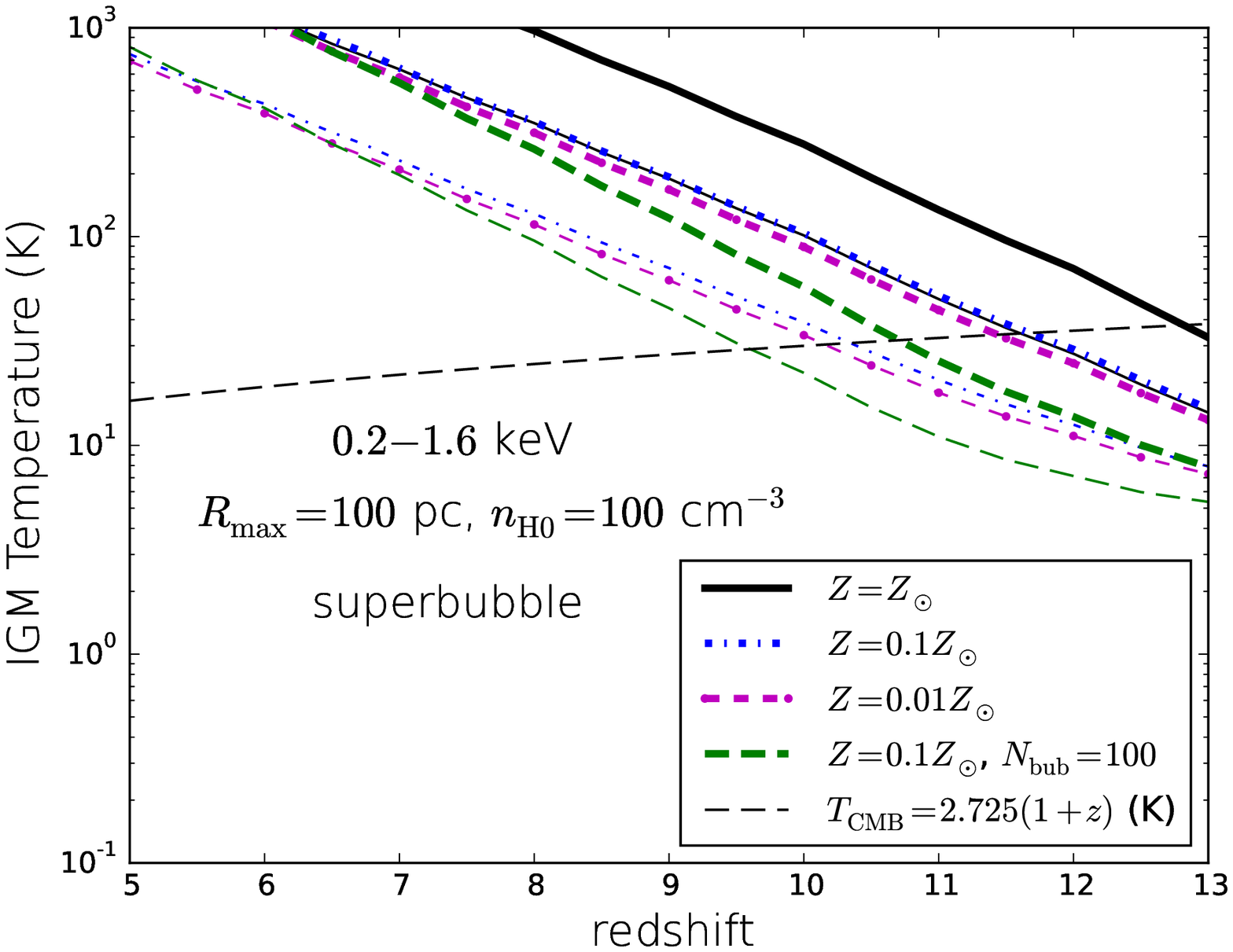}}
\caption{(Left panel) The halo-averaged soft X-ray emissivity as a
  function of redshift for a superbubble expanding to a maxium radius
  $R_{\rm max}=100$~pc into a surrounding medium of hydrogen density
  $n_{\rm H0}=100\,{\rm cm}^{-3}$, for a variety of gas
  metallicities. Also shown is a case in which the star formation rate
  is shared equally among 100 independent superbubbles each reaching a
  maximum radius 100~pc. (Right panel) The evolution in the
  temperature of the still neutral IGM using the X-ray emissivities in
  the left panel. The heavy lines assume all the X-ray energy is
  deposited in the IGM. The light lines allow for losses to secondary
  ionizations, with only 36 percent of the X-ray energy deposited as heat
  (see text). The dashed line rising towards higher redshifts is the
  CMB temperature.
}
\label{fig:TEvolSB}
\end{figure*}

The evolution in the superbubble soft X-ray emissivity in the $0.2-1.6$~keV band is shown in the left panel of Fig.~\ref{fig:TEvolSB}, averaged over halo masses. The evolution is remarkably flat, and generally exceeds the corresponding emissivity for pressure-driven winds, especially at early cosmological times. For metallicities $Z<0.1Z_\odot$, the contribution from emission lines is small.

The enhanced early emissivity compared with the pressure-driven wind model heats the neutral component of the IGM to temperatures above that of the CMB at earlier times. Even allowing for an X-ray heating efficiency factor of 36 percent, the IGM temperature should exceed the CMB temperature by $z\sim9.5-10.5$ for $Z\le0.1Z_\odot$. The temperatures tend to converge for all models at $z<7$.

\subsection{Heating in FiBY simulation}
\label{subsec:heatFiBY}

\begin{figure}
\includegraphics[width=3.3in]{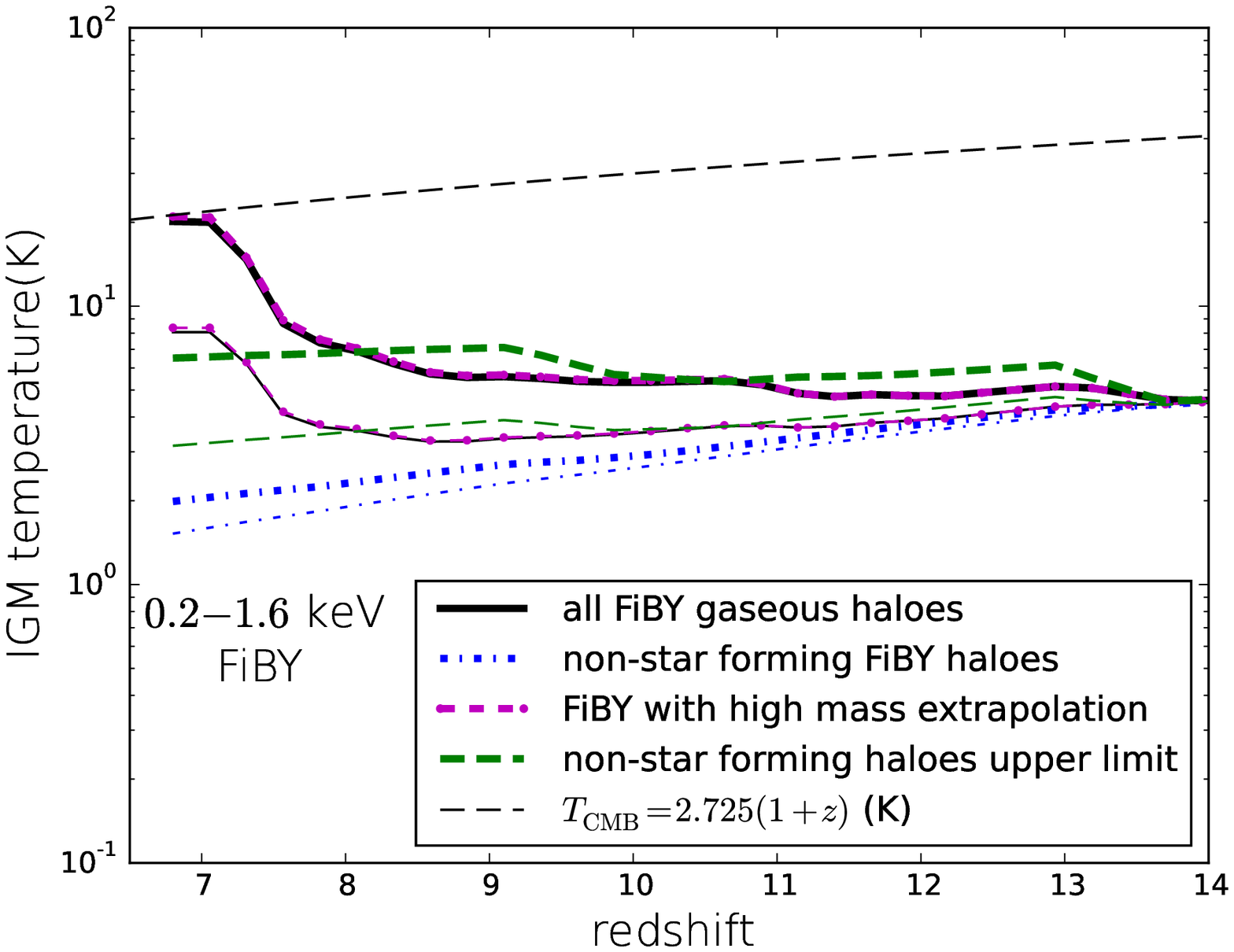}
\caption{The evolution in the temperature of the still neutral IGM
  using the X-ray emissivity from the FiBY simulation, allowing for
  internal soft X-ray attenuation from the galaxies. The heavy lines
  assume all the X-ray energy is deposited in the IGM. The light lines
  allow for losses to secondary ionizations, with only 36 percent of the
  X-ray energy deposited as heat (see text). The dashed line rising
  towards higher redshifts is the CMB temperature. Below $z=9$, the
  temperatures are only formal since the IGM is reionized in the
  simulation by this time.
}
\label{fig:TEvolFiBY}
\end{figure}

The evolution in the IGM temperature including the heat input from
X-rays in the range $0.2-1.6$~keV is shown in Fig.~\ref{fig:TEvolFiBY}
for the FiBY simulation. The solid lines show the heating from all
gaseous haloes in the simulation. Since the box is too small to
capture massive haloes, a correction is estimated by adopting the mean
heating rate per halo mass for the most massive haloes, and applying
this factor to haloes so massive that fewer than 1 is expected in the
simulation volume based on the halo mass function of
\citet{2007MNRAS.374....2R}. The sum of the correction and the
contribution from haloes in the simulation volume (dot-dashed curve)
shows little heat is missing from the simulation. Below $z=9$, the
temperatures are only formal since sufficient numbers of UV
photoionizing photons have escaped the haloes to reionize the IGM by
this time \citep{2013MNRAS.429L..94P, 2015MNRAS.451.2544P}. The IGM
temperature thus never exceeds the CMB temperature prior to
reionization.

Since gas falling into the dark matter haloes will heat to X-ray
temperatures as it flows into the haloes and shocks, non-star forming
haloes may also contribute to the IGM heating
\citep{1997ApJ...475..429M}. The short-dashed-dotted (blue) lines show
that such gaseous haloes in the simulation volume contribute very
little to the heating. Since the gaseous systems make up typically
10--50 per cent of all haloes, depending on mass, a correction factor
is estimated by applying the heating rate per halo mass as a function
of halo mass to all haloes. This provides an upper limit, likely
generous, to the amount of heating that may potentially be missing
from the simulation if the gas has not been adequately resolved in the
gas poor systems. The result is shown by the long-dashed (green)
lines. At $z>9$, before reionization completes, the upper limit is
comparable to the heating rate from all the identified gaseous haloes,
including those forming stars.

\section{Discussion and conclusions}

\begin{figure*}
\scalebox{0.45}{\includegraphics{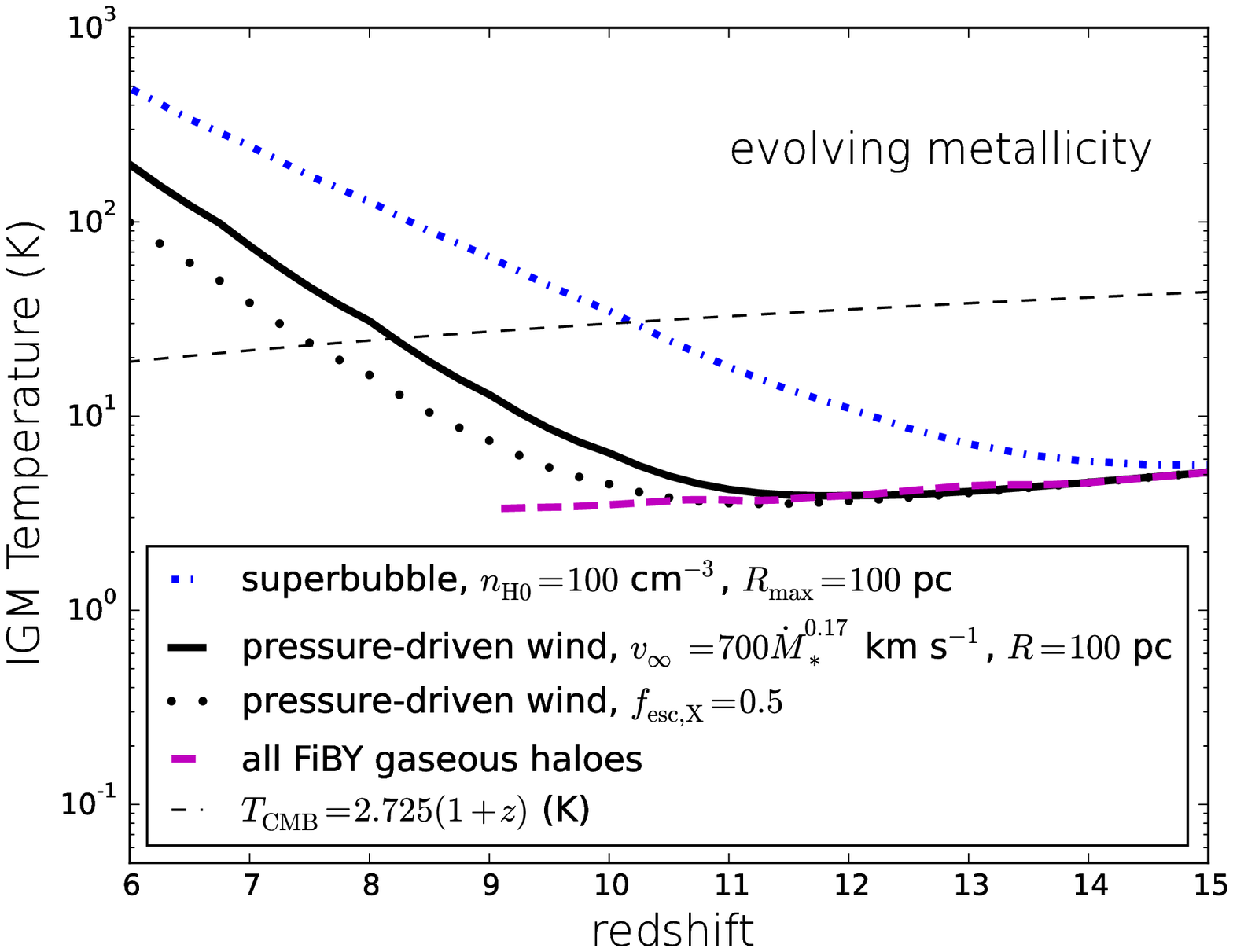}\includegraphics{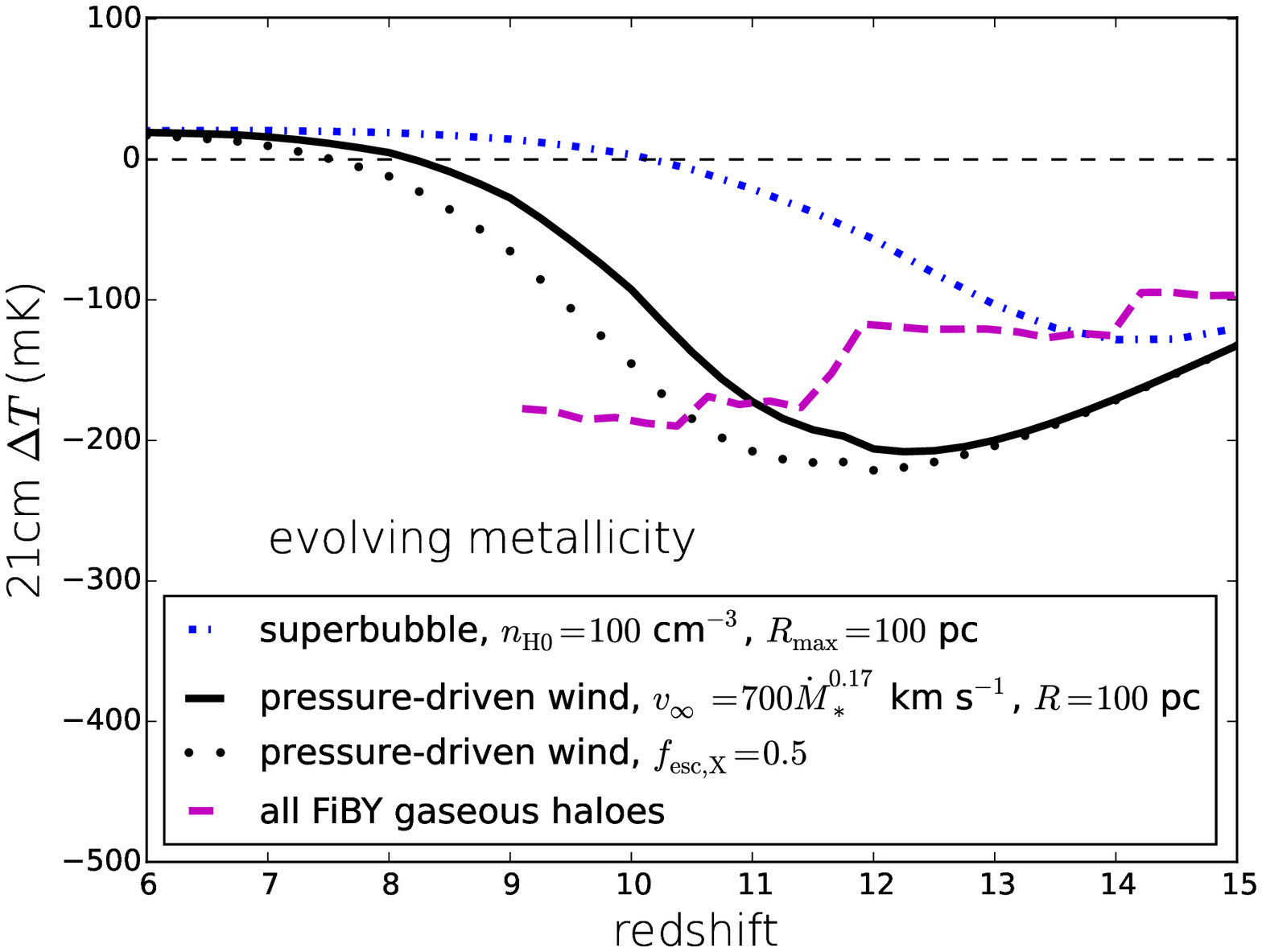}}
\caption{(Left panel) The IGM temperature as a function of redshift
  for pressure-driven wind, expanding superbubble and FiBY simulation,
  allowing for an evolving metallicity. (Right panel) The 21~cm IGM
  differential temperature for the IGM temperatures in the left
  panel. The IGM is reionized for $z<9$ in the FiBY simulation.
  }
\label{fig:TEvolSBWF}
\end{figure*}

\begin{figure}
\includegraphics[width=3.3in]{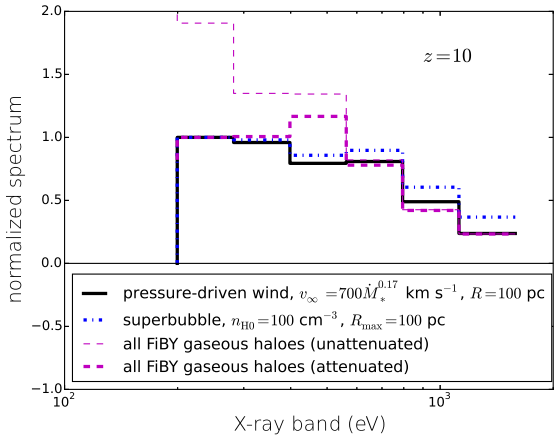}
\caption{Metagalactic X-ray emission spectrum at $z=10$ for the pressure-driven wind model, superbubble model and FiBY galaxies. The spectrum is integrated over energy bands of constant logarithmic width and normalized at the band $200-280$~eV. The FiBY spectrum is shown both without (light dashed line) and with (heavy dashed line) internal galactic absorption.
}
\label{fig:EmissMG}
\end{figure}

\begin{figure*}
\scalebox{0.45}{\includegraphics{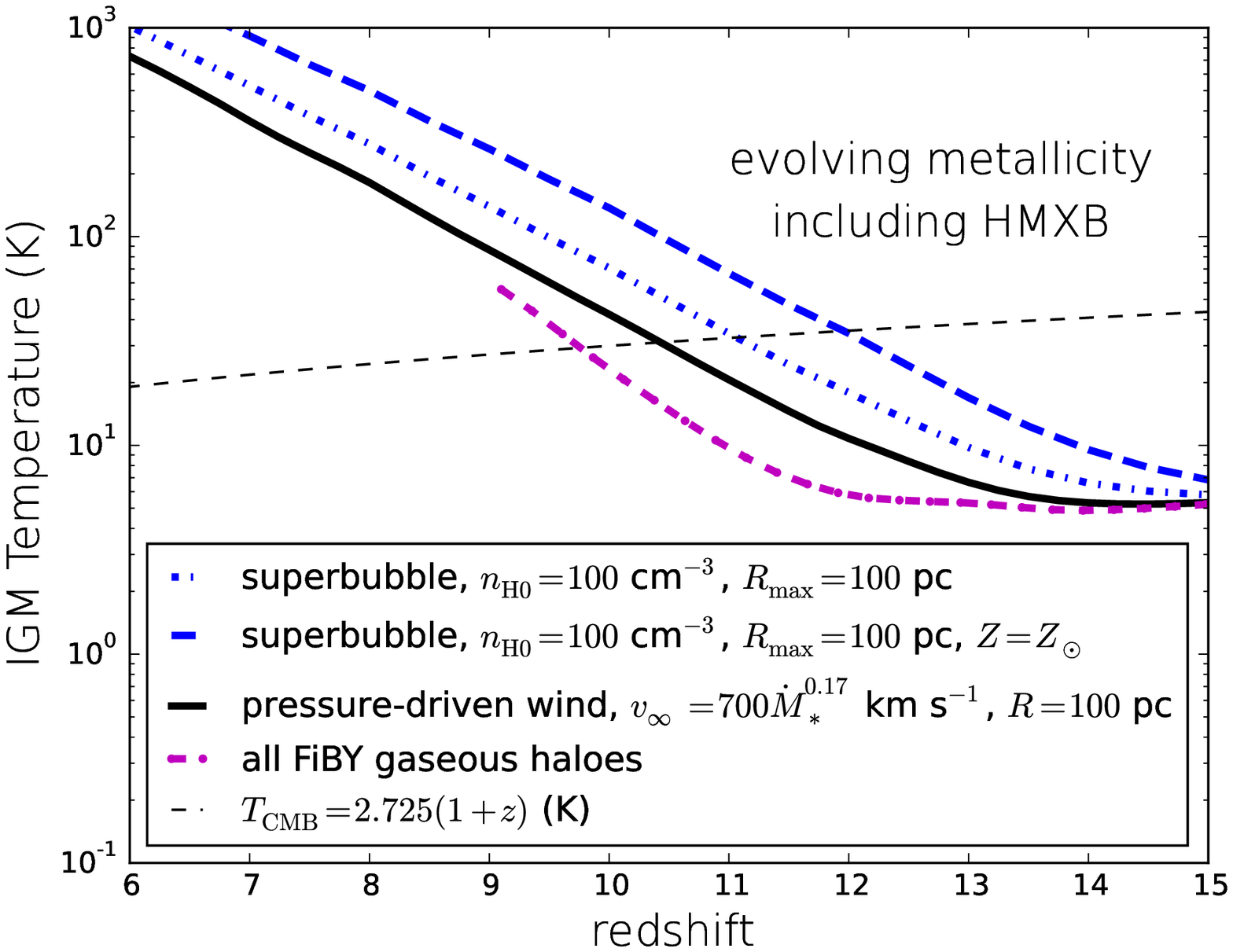}\includegraphics{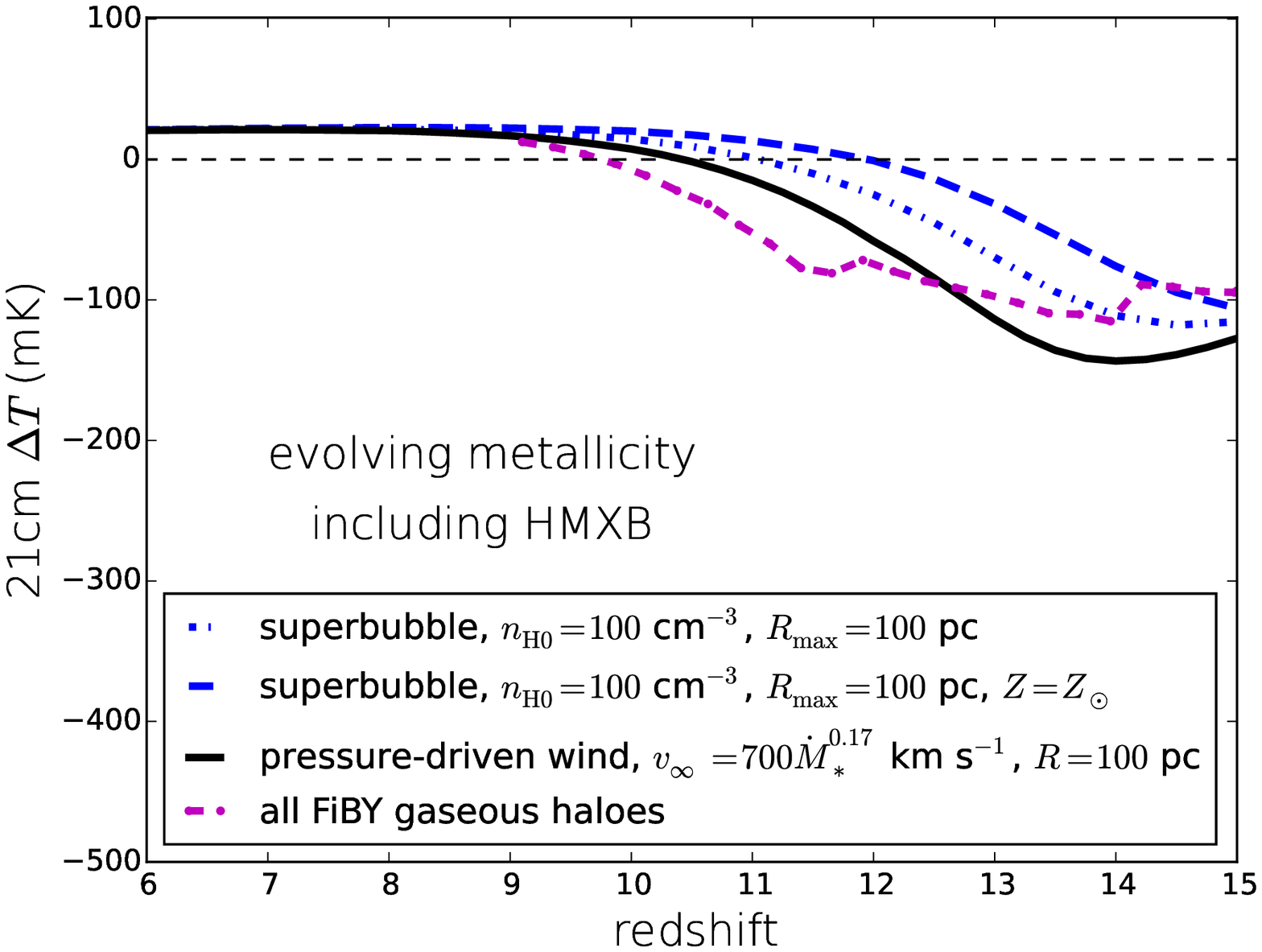}}
\caption{(Left panel) The IGM temperature as a function of redshift
  for pressure-driven wind, expanding superbubble and FiBY simulation,
  allowing for an evolving metallicity and including heating by
  high-mass X-ray binaries. (Right panel) The corresponding 21~cm IGM
  differential temperatures. The IGM is reionized for $z<9$ in the
  FiBY simulation.
}
\label{fig:TEvolSBWF_HMXB}
\end{figure*}

Best-estimate predictions for the evolution of the IGM temperature for
the wind models are provided in the left panel of
Fig.~\ref{fig:TEvolSBWF}. Shown are results for a pressure-driven wind
with source radius $R=100$~pc and $v_\infty=700\dot M_*^{0.17}\,{\rm
  km\, s^{-1}}$, a superbubble expanding to a maxium radius $R_{\rm
  max}=100$~pc into a surrounding medium of hydrogen density $n_{\rm
  H0}=100\,{\rm cm}^{-3}$, and the FiBY simulation. The FiBY results
are truncated at $z<9$ since the simulation introduces a sufficient
abundance of photoionizing UV photons into the IGM to complete
reionization by this time \citep{2013MNRAS.429L..94P,
  2015MNRAS.451.2544P}. For the pressure-driven wind and superbubble
models, the gas metallicity evolution from \citet{2017ApJ...840...39M}
has been used. A heating efficiency of 36 percent was adopted, allowing for
losses to secondary ionizations, following
\citet{2017ApJ...840...39M}. The IGM temperature is approximated by
adding the time-integrated heat input by the soft X-rays to the
temperature without heating sources, using the fit $T_{\rm IGM,\,
  no\,\,heating}\simeq0.023(1+z)^{1.95}$ \citep{2011MNRAS.417.1480M},
based on integrating the ionization and thermal equations for a gas of
primordial composition after the recombination era with the publicly
available code RECFAST \citep{2000ApJS..128..407S}.

The temperature does not rise above the CMB temperature until
$z\simeq8$ in the pressure-driven wind model. Results are shown for
soft X-ray escape fractions $f_{\rm esc, X}=1$ (solid line) and 0.5
(dotted line), the latter corresponding to the mean found in the FiBY
simulation. For the superbubble model, $f_{\rm esc, X}=1$ is
assumed. The spectral shapes of the metagalactic emissivity for the
models are shown at $z=10$ in Fig.~\ref{fig:EmissMG}, computed by
summing the spectra from individual haloes over the halo mass
function. The FiBY emissivity is shown both with and without
attenuation by internal galactic absorption. The attenuation
corresponds to an effective average \HI\ column density $N_{\rm
  HI}\simeq6\times10^{20}\,{\rm cm}^{-3}$. This does not imply the
X-rays are attenuated through a uniform distribution of hydrogen at
this column density, but represents only an average over the galaxy
population as a whole, still allowing for porous regions of very low
attenuation. A more complete description of attenuation in the FiBY
simulations is provided by \citet{2015MNRAS.451.2544P}. The effect of
attenuation in the FiBY galaxies on the heating of the IGM is
discussed in the Appendix. The pressure-driven wind model and
superbubble model emissivities are nearly the same, with the
superbubble model slightly harder. The FiBY galaxies including
attenuation provide a similar metagalactic emissivity to the analytic
models. The superbubble model produces a relatively more intense X-ray
emissivity compared with the pressure-driven wind model or FiBY
galaxies, heating the IGM to temperatures above that of the CMB at the
relatively earlier epoch $z\simeq10$, preceded by a gradual rise. In
contrast, in the FiBY simulation the gas is never heated above the CMB
before reionization completes at $z=9$ in the model. Comparison with
Fig.~\ref{fig:EmissEoRFiBY} suggests the slow heating is a consequence
of the inefficient generation of soft X-rays compared with the
pressure-driven wind model in haloes with star formation rates above
$0.01\,M_\odot\,{\rm yr}^{-1}$. The need for greater numerical
resolution, however, cannot easily be ruled out. The simulations
require high resolution for the radiative transfer computation, which
can be critical especially for high star formation rates or the more
massive haloes, and they are already pushing the computational
limit. The question merits further investigation.

A radio experiment differencing sky images, confined to narrow
wavebands, of patches of ionized and still neutral IGM hydrogen will
measure a 21~cm brightness temperature differential
\begin{equation}
\Delta T \simeq
(26\,{\rm mK})\left(\frac{1+z}{10}\right)^{1/2}\left[1-\frac{T_{\rm
      CMB}(z)}{T_S(z)}\right],
\label{eq:dT21cm}
\end{equation}
\citep{1997ApJ...475..429M, 2011MNRAS.417.1480M}, where $T_{\rm
  CMB}(z)$ is the temperature of the CMB and $T_S(z)$ the spin
temperature of the neutral hydrogen. The spin temperature will be
decoupled from the CMB temperature by a sufficient supply of
Ly$\alpha$ photons through the Wouthuysen-Field mechanism, and
collisional de-excitation between hydrogen atoms. It is given by
\begin{equation}
T_S = \frac{T_{\rm CMB} + y_\alpha T_\alpha + y_cT_K}{1+y_\alpha+y_c},
\label{eq:TSpin}
\end{equation}
where $T_K$ is the kinetic temperature of the gas, $T_\alpha$ is the
\lq\lq light temperature\rq\rq of the Ly$\alpha$ photons, with weight
factors $y_\alpha$ and $y_c$ depending on the Ly$\alpha$ and
collisional de-excitation rates \citep{1959ApJ...129..536F,
  1997ApJ...475..429M}. In IGM conditions, the light temperature
rapidly relaxes to the kinetic temperature of the gas
\citep{2006MNRAS.370.2025M}. Continuum photons emitted between the
Ly$\alpha$ and Ly$\beta$ transitions will redshift into the local
Ly$\alpha$ line, where they will scatter many times (given in number
at large distances from galaxies by the inverse of the local Sobolev
parameter) before redshifting away \citep{1959ApJ...129..536F,
  2012MNRAS.426.2380H}. The Starburst99 model \citep{Leitherer99} for
continuous star formation following a Salpeter stellar initial mass
function between $0.1-100\,M_\odot$ predicts a production rate of
photons between Ly$\alpha$ and Ly$\beta$ of
$1.5\times10^{53}\,{\rm s}^{-1}$ for a metallicity of $Z=0.05Z_\odot$
after $10^7$~yr, at which time it plateaus. Following
\citet{1997ApJ...475..429M}, this corresponds to
$y_\alpha\simeq2000S_\alpha T_{\rm K}^{-1}{\dot\rho_*} (1+z)^{3/2}$,
where $\dot\rho_*$ is the (comoving) star formation rate in
$M_\odot\,{\rm yr}^{-1}\,{\rm Mpc}^{-3}$, and $S_\alpha\simeq0.8$
accounts for the shape of the line profile \citep{2004ApJ...602....1C,
  2006MNRAS.367..259H}. Collisional de-excitations, also included,
contribute only a small correction over the redshifts of interest.

The right panel of Fig.~\ref{fig:TEvolSBWF} shows the evolution of the
resulting 21~cm differential brightness temperature corresponding to
the models in the left panel. For the pressure-driven wind model,
X-ray heating only becomes influential at $z\lta12.5$, with the 21~cm
signature making the transition from absorption to emission at $z_{\rm
  trans}\simeq8.2$ (observed 21~cm frequency 155~MHz). A minimum in
the absorption signature occurs at $z\simeq12.2$ (110~MHz), with
$\Delta T\simeq-210$~mK. If reionization has not completed by $z=8.4$,
the absorption corresponds to an IGM temperature of $T_{\rm
  IGM}\simeq21$~K, only a few times the lower limit set by PAPER-64
measurements \citep{2015ApJ...809...62P} of $T_{\rm IGM}>5$~K at
$z=8.4$, inferred at 95 percent confidence if the neutral fraction is
between 10--85 percent. Adopting $f_{\rm esc, X}=0.5$ decreases the
redshift of the transition to emission to $z_{\rm trans}\simeq7.5$,
with $T_{\rm IGM}\simeq11$~K at $z=8.4$.

In the FiBY simulation, the transition to emission does not take place
before reionization completes, with a minimum absorption temperature
differential of $\Delta T\simeq-190$~mK at $z\simeq10.4$ (125~MHz). At
$z>11$, the absorption signal is shallower than for the
pressure-driven wind model, even though the gas temperatures are
nearly the same. This is a reflection of the relatively lower star
formation rates in the FiBY simulation compared with that of
\citet{2015ApJ...813...21M} adopted for the pressure-driven wind
model. The production of Ly$\alpha$ to Ly$\beta$ continuum photons is
consequently lower in the FiBY simulation, with weaker decoupling of
the spin temperature from the CMB. By $z<11$, the spin temperature is
well-coupled to the gas kinetic temperature for both the
pressure-driven model and the FiBY simulation, however the
pressure-driven wind model has warmed the gas to higher temperatures
than the FiBY galaxies have done. As a result the FiBY signal remains
more strongly in absorption until reionization completes.

The superbubble model results in earlier heating than either the
pressure-driven wind model or the FiBY galaxies provide, with a
transition redshift to emission at $z_{\rm trans}\simeq10.3$
(125~MHz). A minimum occurs in the absorption differential temperature
rather earlier than in the other models, reaching $\Delta
T\simeq-128$~mK at $z\simeq14$ (95~MHz).

High-mass X-ray binaries in the galaxies are expected to substantially
boost the IGM heating rate. Their contribution has been estimated by
\citet{2017ApJ...840...39M} from population synthesis models. Because
the star formation rates here differ from that used in their estimate,
we have rescaled their heating rates by the ratio of star formation
rates, using the fiducial star formation rate in
\citet{2017ApJ...840...39M} (and normalized to a Sapleter stellar
initial mass function). As shown in the Appendix, this shifts the
redshift at which the IGM temperature exceeds the CMB temperature to
$z_{\rm trans}\simeq9.5-10.5$ from high mass X-ray binary heating alone.

Fig.~\ref{fig:TEvolSBWF_HMXB} shows the evolution of the IGM
temperature and differential 21~cm brightness temperature for the
combined heating from galactic winds and HMXBs. The crossing point of
the IGM and CMB temperatures occurs at $z_{\rm trans}\simeq9.5-11$
($120-130$~MHz), except for the extreme superbubble model with solar
metallicity, for which the transition is at the somewhat earlier epoch
$z_{\rm trans}\simeq12$ (110~MHz). The differential brightness
temperatures all show distinctive patterns with redshift (or
frequency). The deepest minimum is produced by the pressure-driven
wind model, with a minimum 21~cm brightness temperature differential
of $\Delta T\simeq-140$ at $z\simeq14$ (95~MHz). The FiBY model
produces a somewhat shallower minimum of $\Delta T\simeq-115$~mK, also
at $z\simeq14$. The highest redshift minima are for the superbubble
models, with $\Delta T\simeq-120$~mK at $z\simeq14.5$ (90~MHz) for the
model with evolving metallicity, and $\Delta T\simeq-110$~mK at
$z\simeq15.5$ (85~MHz) for the extreme model with solar
metallicity. While the earlier minimum absorption redshifts and
earlier transitions from absorption to emission may provide a means of
indicating the presence of HMXB heating, it is noted the shift to
earlier redshifts is model dependent:\ the shift is only moderate for
the superbubble model.

\citet{2017ApJ...840...39M} adopted an amount of attenuation
corresponding to an average \HI\ column density of
$3\times10^{21}\,{\rm cm}^{-2}$, as indicated by observations of nearby
star-forming galaxies. Had they adopted the smaller amount of
attenuation found in the FiBY simulation for the X-ray gas produced by
winds, the amount of HMXB heating would increase. The relative amounts
of attenuation of X-rays from the HMXB population and the extended
X-ray bubbles produced by pressure-driven winds in young galaxies is
unknown. The simulations analyzed by \citet{2017MNRAS.469.1166D}
suggests a wide spread in the amount of attenuation of X-rays from
HMXBs, resulting in peak emission at lower X-ray energies than in the
fiducial model of \citet{2017ApJ...840...39M}. The mean \HI\ column
density they obtain is still a factor of a few higher than the
effective column density found for the wind-generated X-rays in the
FiBY galaxies.

For the superbubble model, the X-ray emission increases with the local
gas density, while too high an amount of mass loading in a
pressure-driven wind could lower the gas temperature and substantially
suppress the X-ray production rate. If local extinction in the regions
where HMXBs form greatly reduces the HMXB heating contribution, the
transition from absorption to emission may occur as late as $z_{\rm
  trans}<8.5$, if at all, in the models considered. When heating by
galactic winds is combined with HMXB heating, the models disfavour a
global 21~cm emission signal at $z>12$. An early emission signal would
then suggest new astrophysics. Possibilities include substantially
more efficient HMXB formation than in the local Universe, a different
wind environment or alternative wind production mechanisms, or
additional sources of heating altogether. Other heating sources that
have been considered include high mass X-ray binaries formed from
Population III stars \citep{2016ApJ...832L...5X} and an early
population of mini-QSOs \citep{2006ApJ...637L...1K} or galaxies
harbouring moderately massive black holes
\citep{2015MNRAS.454.3771P}. An initial mass function tilted towards
massive Pop~III stars could also lead to greater heating from
pair-instability supernovae remnants.

While average global heating and 21~cm signals were computed for the
models, the FiBY simulation shows that on small scales the rate of
soft X-ray heating can vary widely for two reasons, anisotropic soft
X-ray attenuation and a varying star-formation rate. Soft X-ray
attenuation is especially strong in the more massive haloes (above
$10^8\,M_\odot$), and its average can vary considerably from halo to
halo, or in direction for a single halo. For halo masses
$10^6-10^9\,M_\odot$, star formation is stochastic, with as little as
10 percent of the lower mass haloes actively producing stars. In the
early stages of reionization, both the gas heating and decoupling of
the spin temperature from the CMB may consequently show large
variations on small spatial scales. High spatial frequency modes of
the 21~cm power spectrum may therefore provide an additional means of
testing star-formation and feedback scenarios, a topic worthy of
further investigation.

\appendix

\section{Sensitivity to model assumptions}

\subsection{Pressure-driven wind}

\begin{figure*}
\scalebox{0.45}{\includegraphics{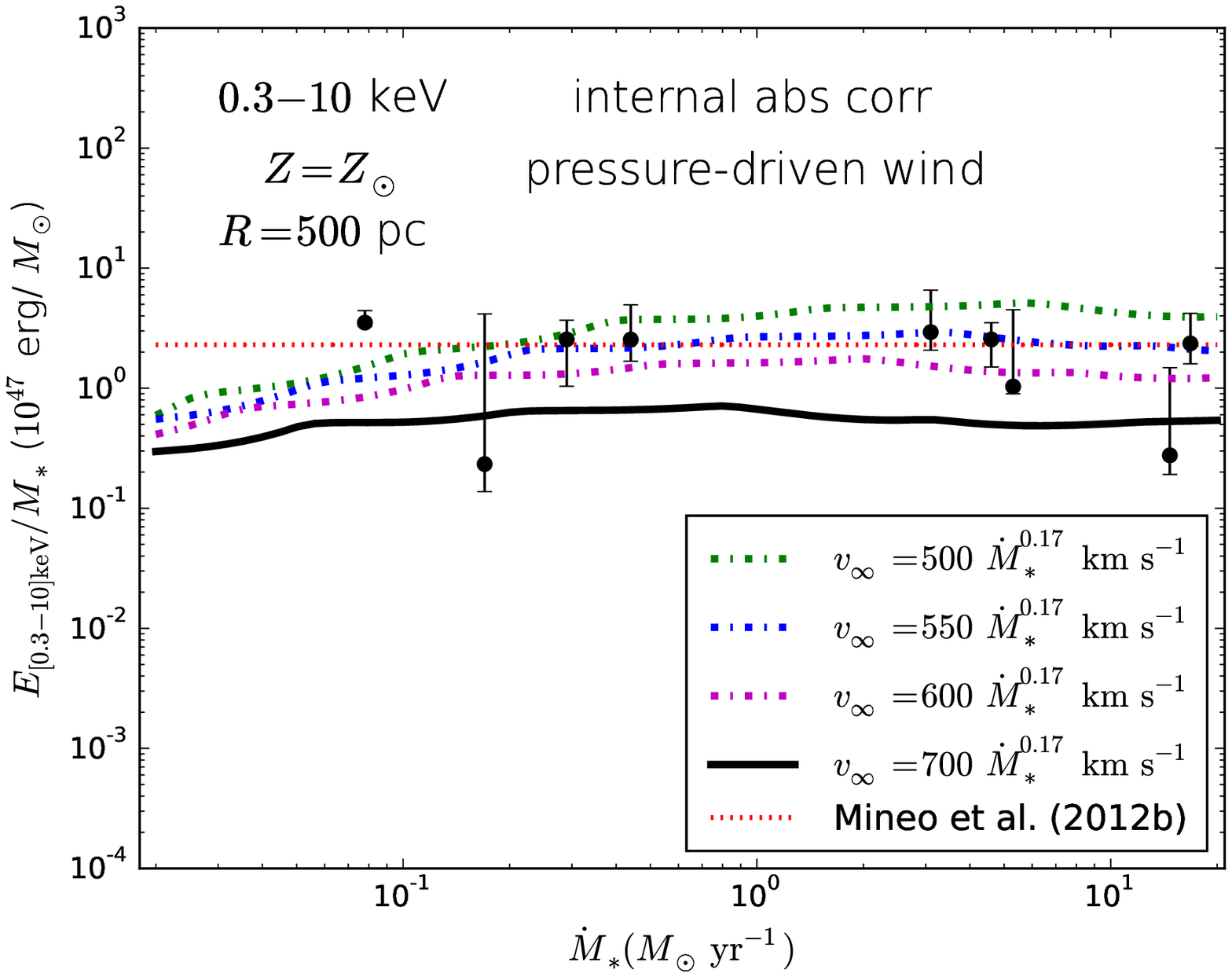}\includegraphics{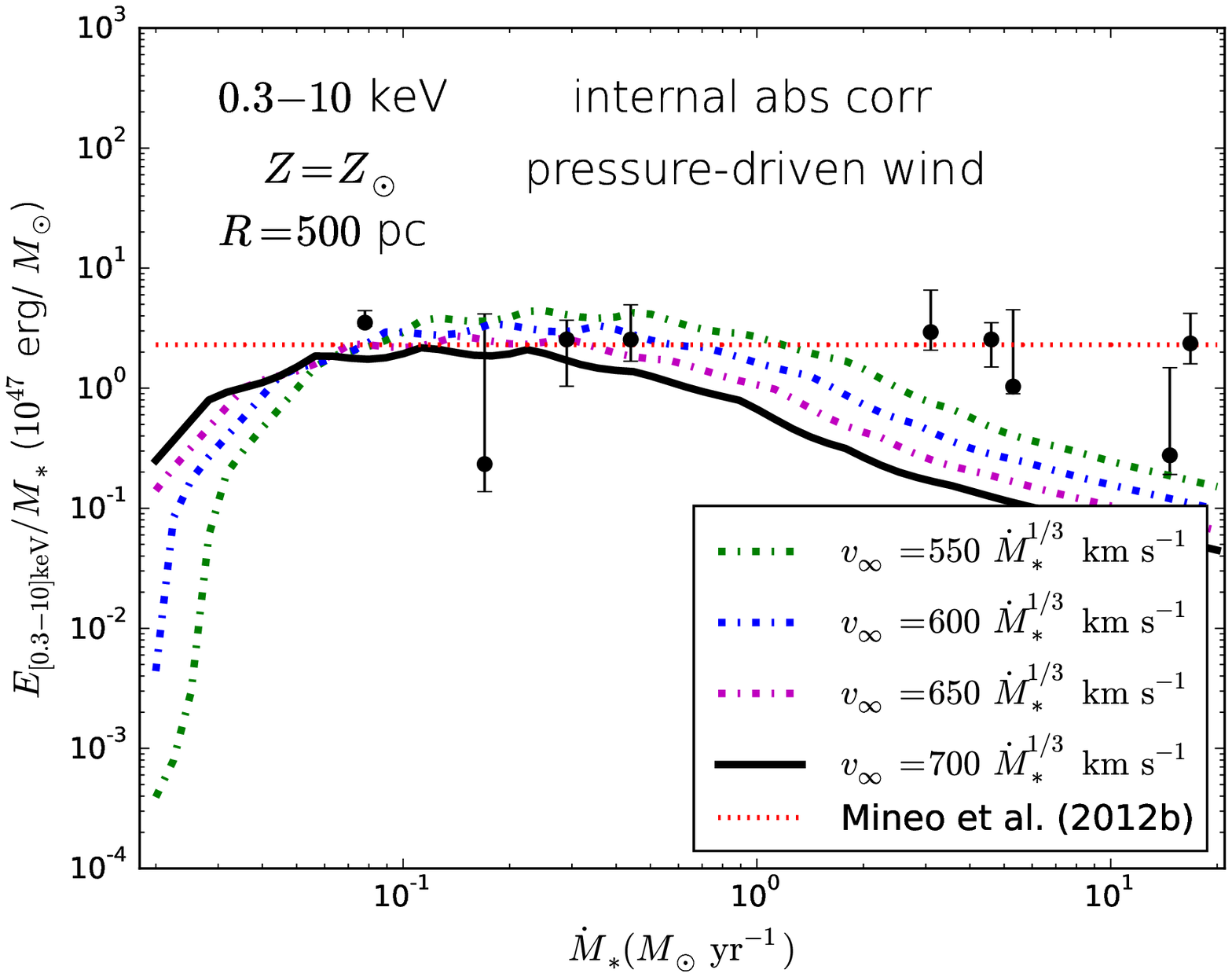}}
\caption{X-ray emissivity per solar mass of stars formed within a
  region of radius $R=500$~pc by a pressure-driven wind with
  asymptotic wind velocity $v_\infty=V_1{\dot M}_*^\alpha$, where
  $\alpha=1/6$ in the left panel and $1/3$ in the right panel. Shown
  for X-ray energies integrated over $0.3-10$~keV for solar
  metallicity. The data points are from \citet{2012MNRAS.426.1870M},
  and include corrections for internal galactic absorption.
}
\label{fig:vinf17}
\end{figure*}

The mass loading is related to the asymptotic wind velocity $v_\infty$
for a pressure-driven wind according to Eq.~(\ref{eq:eta}). Motivated
by the measured correlation of the diffuse X-ray luminosity with the
star formation rate, a power-law relation of the form
$v_\infty=V_1{\dot M}_*^\alpha$ has been assumed. An inherent
ambiguity in the X-ray data arises from an uncertain correction for
internal absorption within the galaxies. Taking the results of
\citet{2012MNRAS.426.1870M} for the $0.3-10$~keV band for galaxies
with spectral evidence for absorption, the values $V_1\simeq700\,{\rm
  km\,s^{-1}}$ and $\alpha\simeq1/6$ are found to provide a good fit
to the data for a star forming region of radius $R=100$~pc and solar
metallicity \citep{2016MNRAS.461.2762M}. For a larger star forming
region of 500~pc radius, $V_1\simeq550\,{\rm km\,s^{-1}}$ is required,
as shown in the left panel of Fig.~\ref{fig:vinf17}. As shown in the
right panel, adopting $\alpha=1/3$, as suggested by the gas outflow
velocities of \citet{2015ApJ...809..147H}, cannot match the diffuse
X-ray emissivity for both low and high star-formation rates
simultaneously.

\subsection{Superbubble}

\begin{figure*}
\scalebox{0.45}{\includegraphics{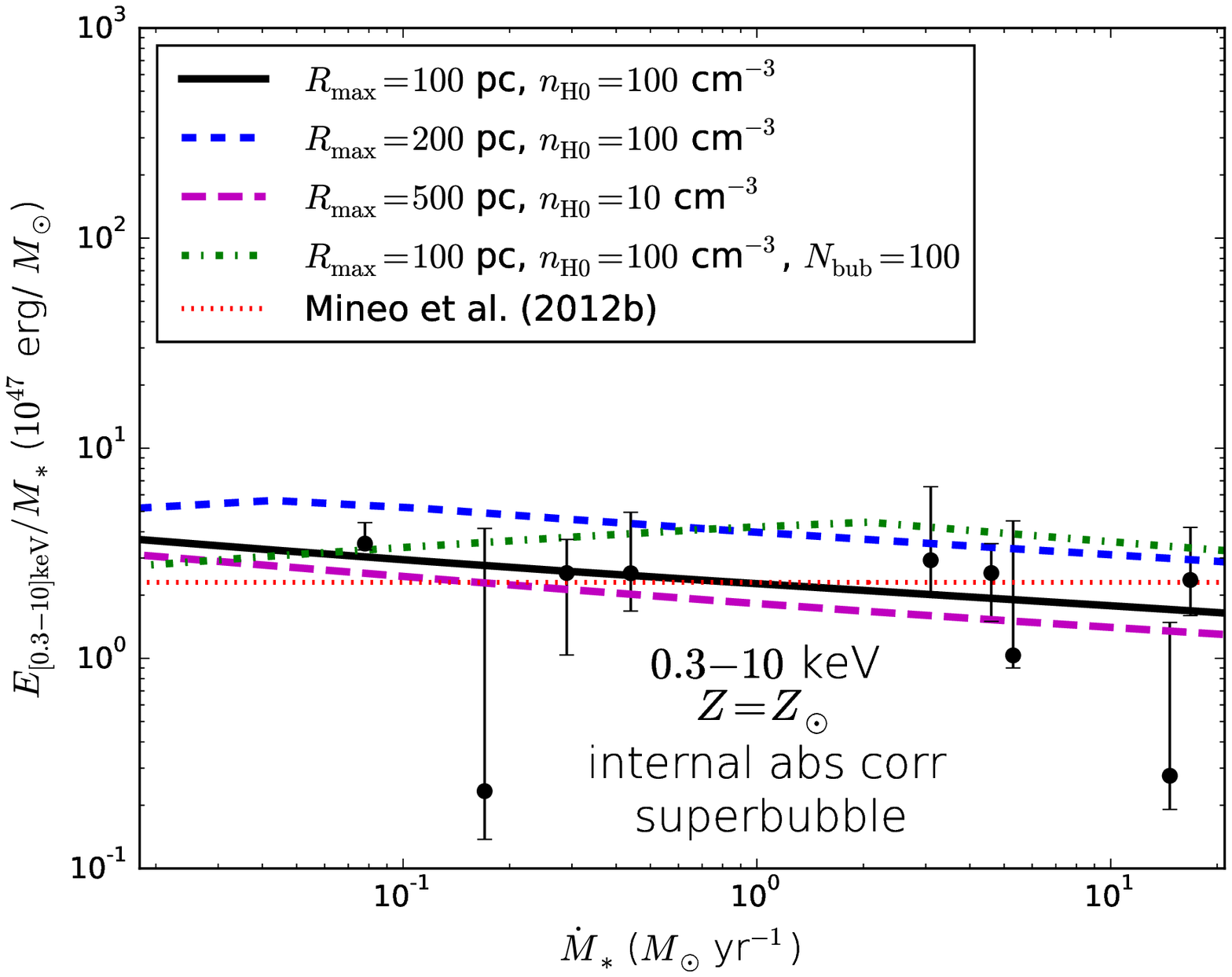}\includegraphics{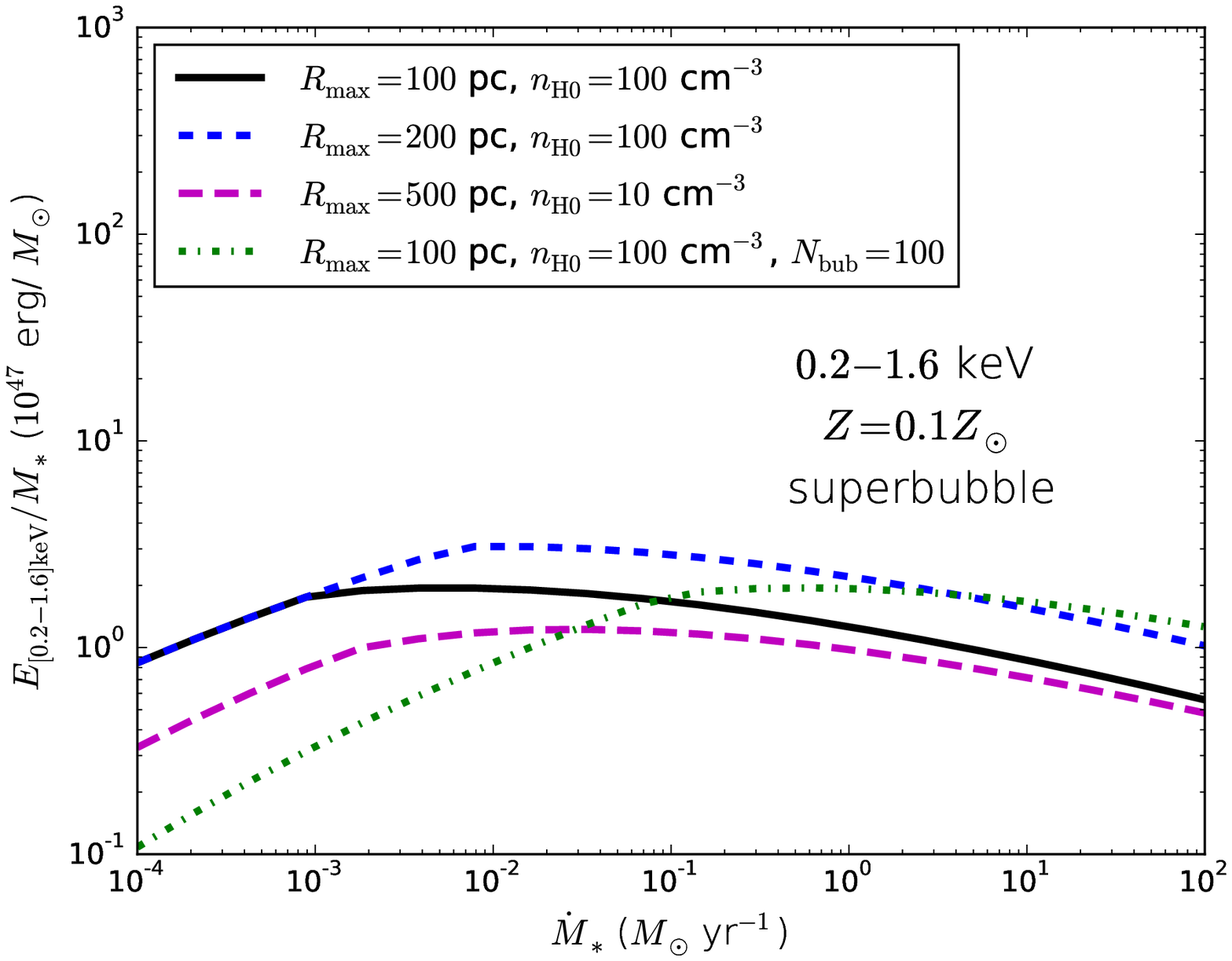}}
\caption{(left panel) X-ray emissivity per solar mass of stars formed from a
  superbubble expanding to a maximum radius $R_{\rm max}$ in a medium
  of ambient hydrogen density $n_{\rm H0}$. Also shown is a case for
  which the star formation rate is shared equally between 100 separate
  superbubbles, each reaching a maximum radius 100~pc. Shown for X-ray
  energies integrated over $0.3-10$~keV for solar metallicity. The
  data points are from \citet{2012MNRAS.426.1870M}, and include
  corrections for internal galactic absorption. (right panel) The corresponding X-ray emissivity per solar mass of stars formed integrated over $0.2-1.6$~keV for metallicity $Z=0.1Z_\odot$.
}
\label{fig:SB}
\end{figure*}

Various superbubble models, with varying maximum radius and ambient gas density, are able to recover the X-ray energy emitted per solar mass of stars formed in the energy band $0.3-10$~keV, as shown in the left panel of Fig.~\ref{fig:SB}. One case allows for the indicated star formation rate to be shared equally by $N_{\rm bub}=100$ distinct superbubbles. As shown in the right panel, models normalized to match the $0.3-10$~keV emissivity predict similar X-ray emissivities in the $0.2-1.6$~keV energy band, representative of the X-ray photons absorbed by the IGM during the reionization epoch. Here a more typical metallicity of $0.1Z_\odot$ is adopted. In the case of multiple bubbles, the predicted emissivity declines somewhat more sharply towards low star formation rates.

\subsection{Low intrinsic diffuse X-ray emission}

\begin{figure*}
\scalebox{0.45}{\includegraphics{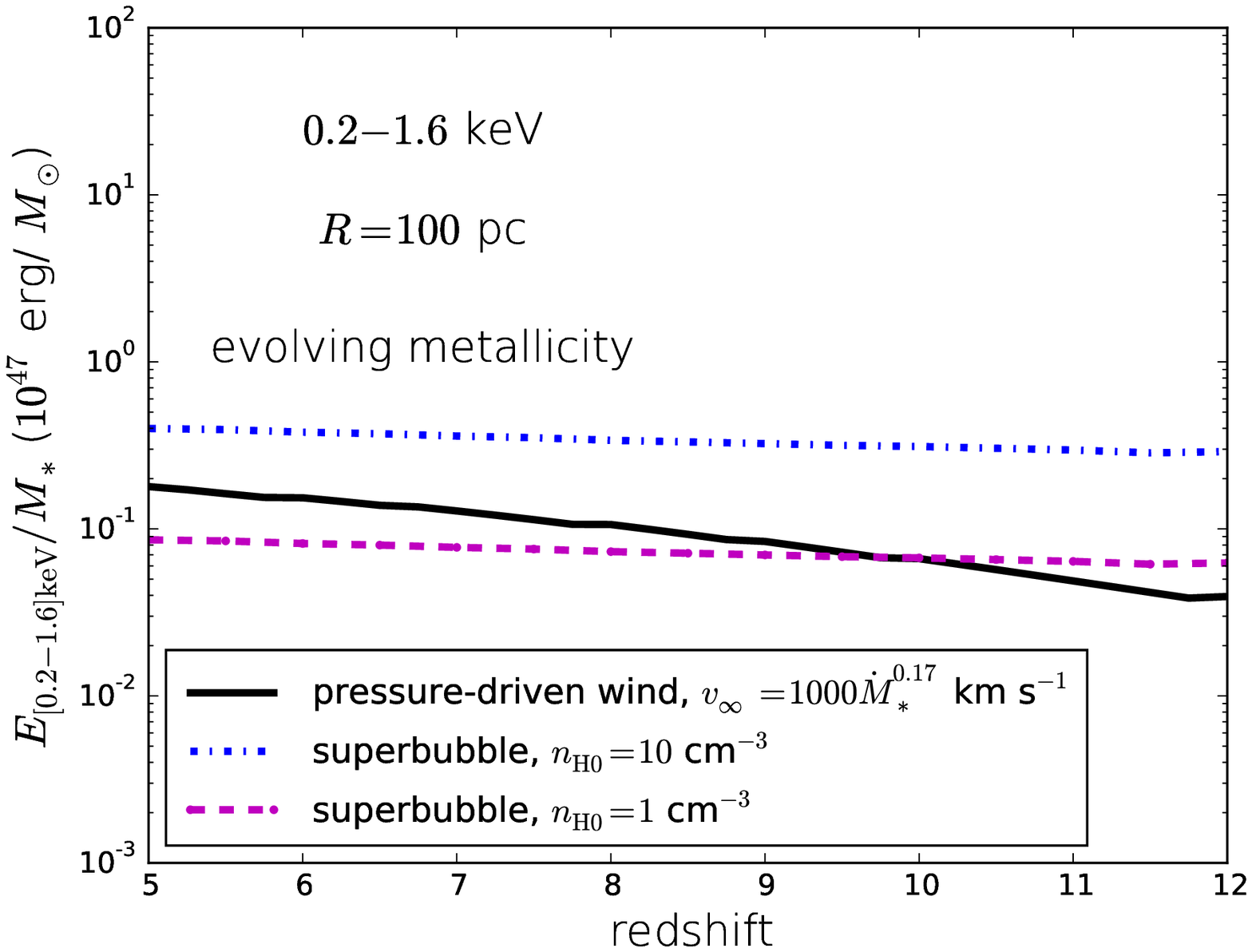}\includegraphics{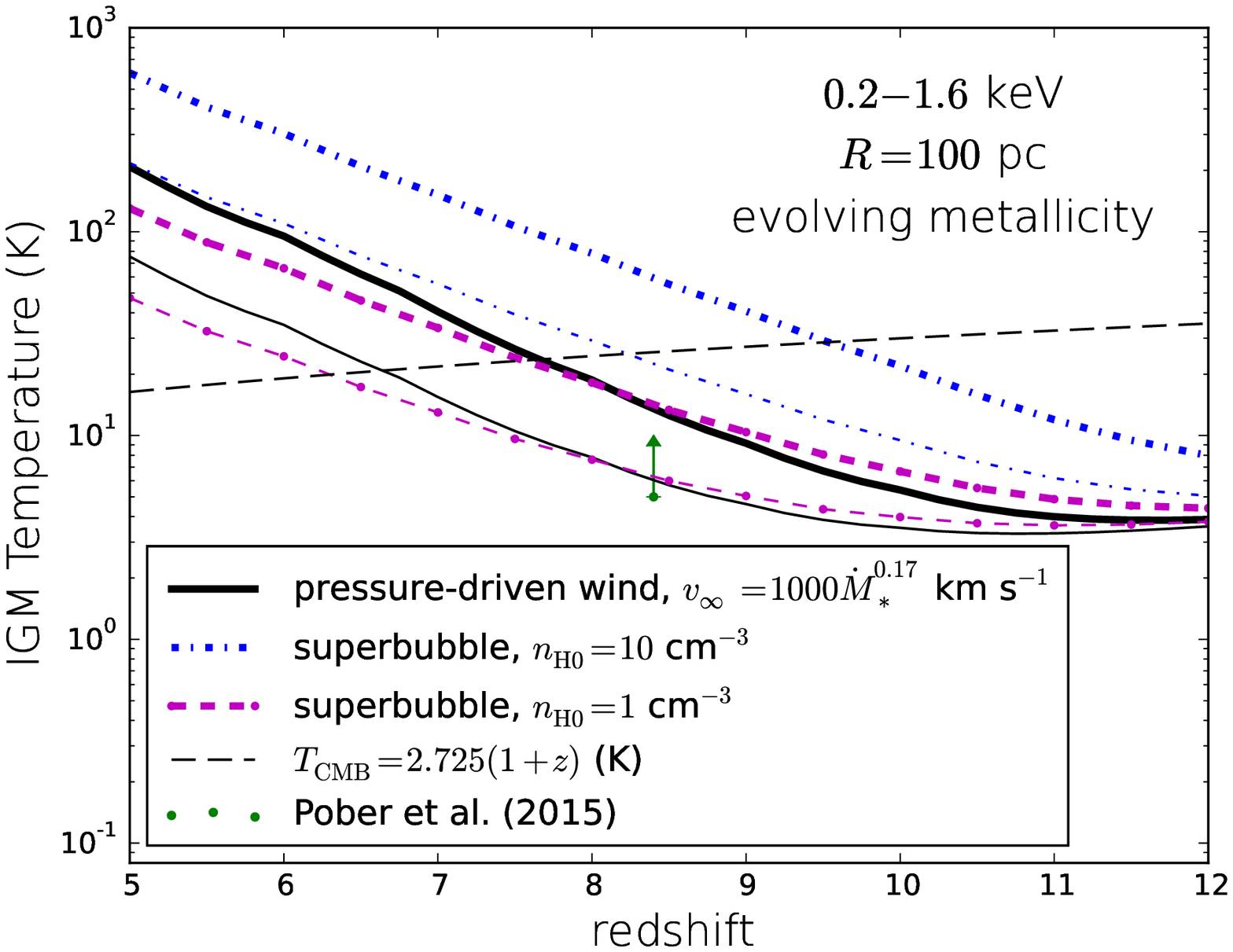}}
\caption{(left panel) X-ray emissivity between $0.2-1.6$~keV per solar mass of stars formed, normalized to a low intrinsic emissivity in nearby galaxies. Shown for a pressure-driven wind with asymptotic wind velocity $v_\infty=1000 {\dot M}_*^{0.17}\,{\rm km\,s^{-1}}$, for which stars form within a region of radius $R=100$~pc (black solid lines), and superbubbles reaching a maximum radius 100~pc expanding into a surrounding medium of density $n_{\rm H0} = 10\,{\rm cm}^{-3}$ (dot-dashed blue lines) and $1\,{\rm cm}^{-3}$ (dashed magenta lines). An evolving gas metallicity is assmed. (right panel) The corresponding evolution of the IGM temperature. The heavy lines assume all the X-ray energy between $0.2-1.6$~keV is used to heat the gas, while the light lines assume a heating efficiency of 36 percent (see text). The data point is the lower temperature limit from PAPER-64 \citep{2015ApJ...809...62P}.
}
\label{fig:lowEmiss}
\end{figure*}

Fig.~\ref{fig:lowEmiss} shows the evolution of the soft diffuse X-ray
emission and IGM temperature for the pressure-driven wind and
superbubble models, normalized to a low intrinsic diffuse X-ray
emissivity between $0.5-2$~keV in nearby galaxies of
$\sim2\times10^{46}$~erg per solar mass of stars formed
\citep{2012MNRAS.426.1870M}, from the models in
\citet{2016MNRAS.461.2762M}. For the pressure-driven wind model, the
X-ray emissivity is best reproduced by correlating the asymptotic wind
velocity with the star-formation rate according to $v_\infty\simeq1000
{\dot M}_*^{0.17}\,{\rm km\,s}^{-1}$, for the star-formation rate
measured in $M_\odot\,{\rm yr}^{-1}$. The superbubble model with
$R_{\rm max}=100$~pc best fits the X-ray data for an ambient total
hydrogen density in the range $1<n_{\rm H0}<10\,{\rm cm}^{-3}$. The
heating rate is reduced by a factor of a few to several compared with
the larger normalization assumed in the rest of this paper, based on
an allowance for internal absorption in the nearby galaxy sample. An
evolving metallicity for the gas is assumed, following
\citet{2017ApJ...840...39M}, who also find that at most 36 percent of
the absorbed soft X-ray energy heats the still neutral IGM, the
remainder being spent on secondary ionizations. Including this
correction, heating of the neutral IGM to temperatures above that of
the CMB is delayed until $z<8$, and possibly as late as $z<6.5$. If
reionization is not substantially completed until $z<8$, as in
\citet{2015ApJ...813...21M}, then the pressure-driven wind model and
the superbubble model for $n_{\rm H0}=1\,{\rm cm}^{-3}$, both with
$T_{\rm IGM}=6$~K at $z=8.4$ allowing for a heating efficiency of 36
percent, are only marginally consistent with the PAPER-64 constraint
on the neutral component temperature $T_{\rm IGM}>5$~K at $z=8.4$
\citep{2015ApJ...809...62P}.

\subsection{Soft X-ray attenuation in FiBY simulation}

\begin{figure}
\includegraphics[width=3.3in]{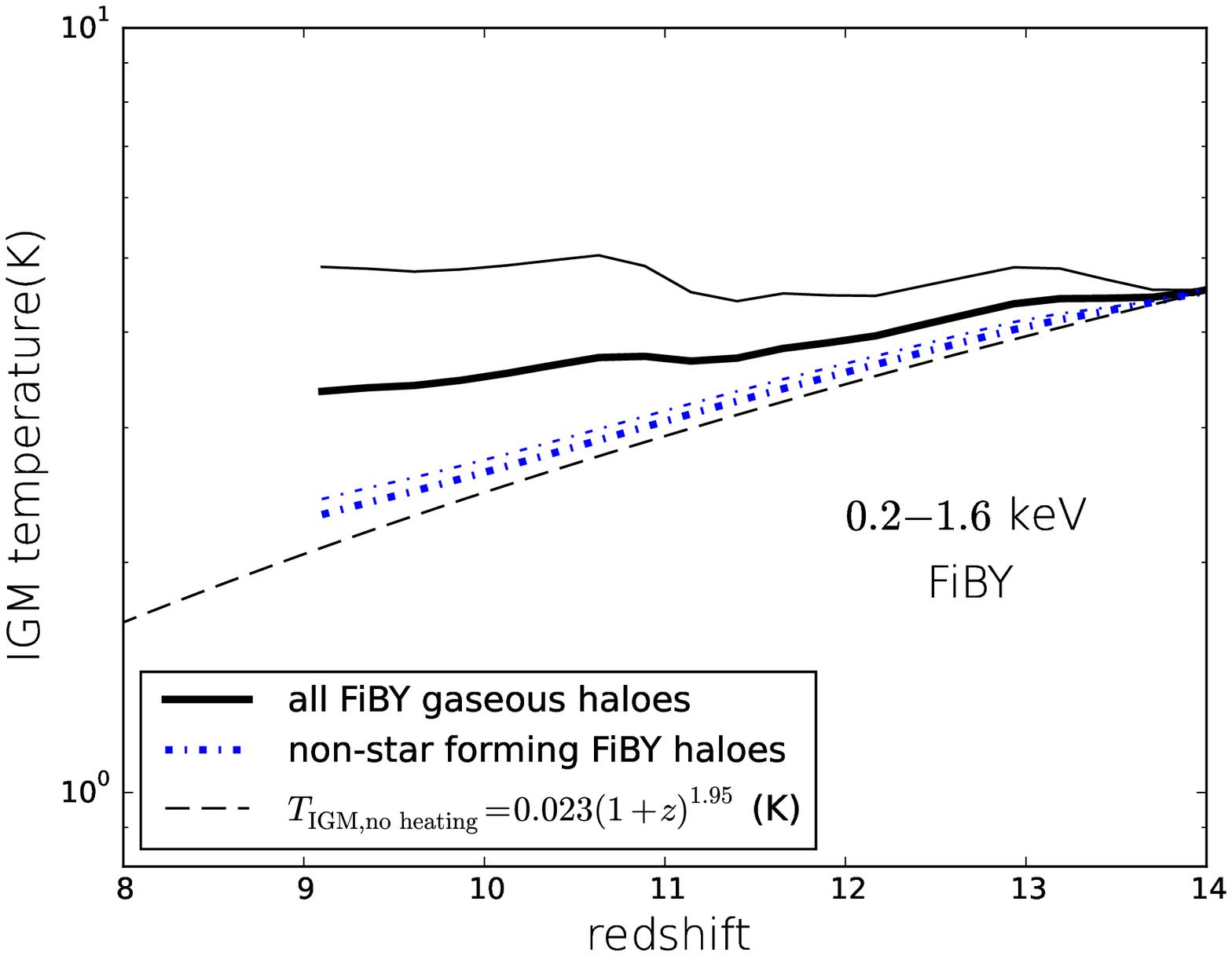}
\caption{The evolution in the temperature of the still neutral IGM
  using the X-ray emissivity from the FiBY simulation. The heavy lines
  allow for internal soft X-ray attenuation from the galaxies, while
  the light lines include no attennuation correction. All curves allow
  for losses to secondary ionizations, with only 36 percent of the X-ray
  energy deposited as heat (see text). The dashed line rising towards
  higher redshifts is the IGM temperature without reheating. The IGM
  is reionized by $z=9$ in the simulation.
}
\label{fig:TEvolFiBY_avna}
\end{figure}

Attenuation by interstellar gas within galaxies in the FiBY simulation
is found to reduce the amount of soft X-rays escaping the haloes by
about 45--55 percent over the redshift range $14>z>9$. The effect of
soft X-ray attenuation on the IGM temperature is shown in
Fig.~\ref{fig:TEvolFiBY_avna}. The heavy lines include attenuation
while the light do not. The solid lines show the temperature evolution
for heating from all gaseous haloes, while the dot-dashed lines show
the temperature evolution from only the non-star forming haloes.

\subsection{High mass X-ray binary heating}

\begin{figure*}
\scalebox{0.45}{\includegraphics{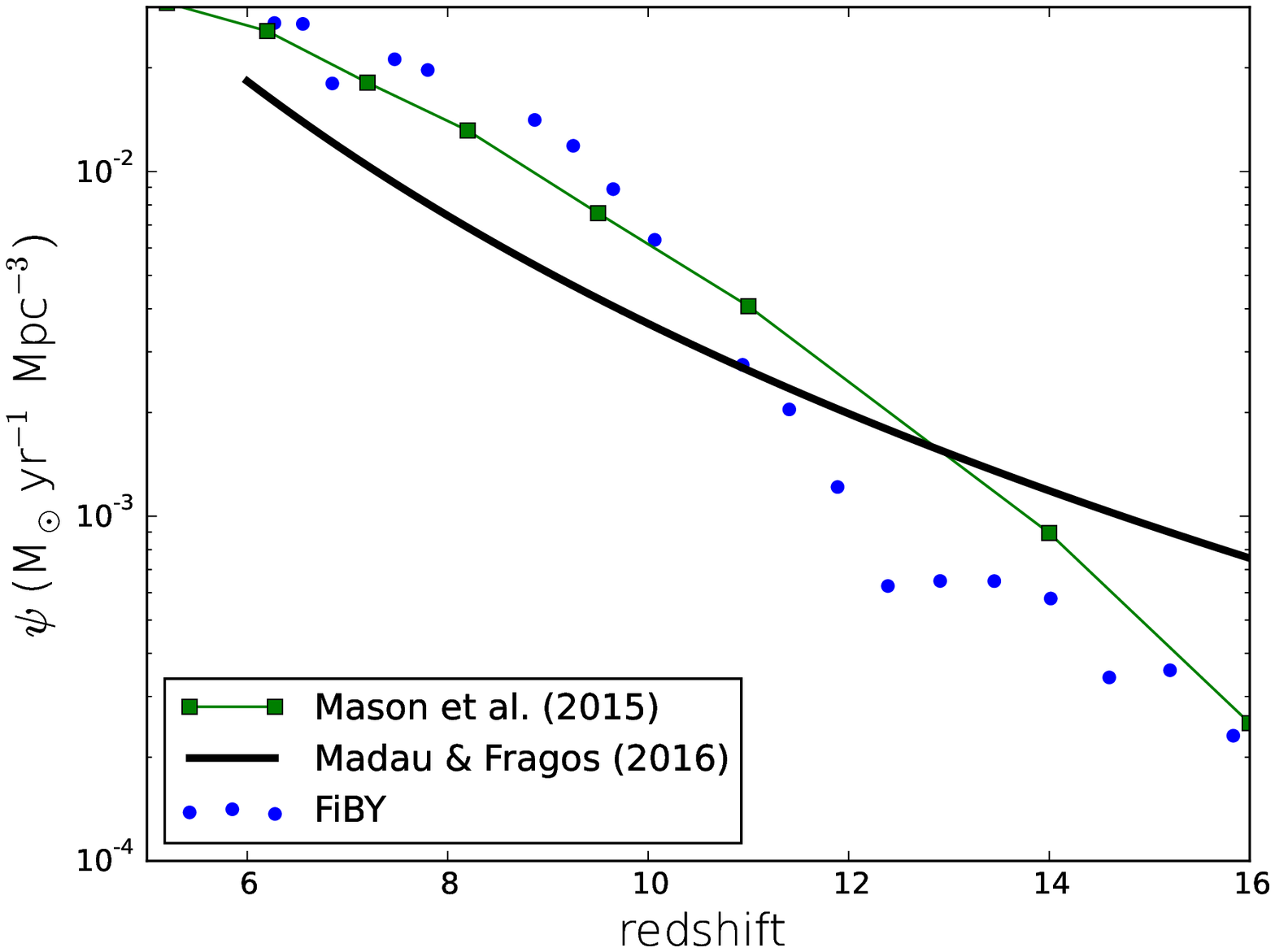}\includegraphics{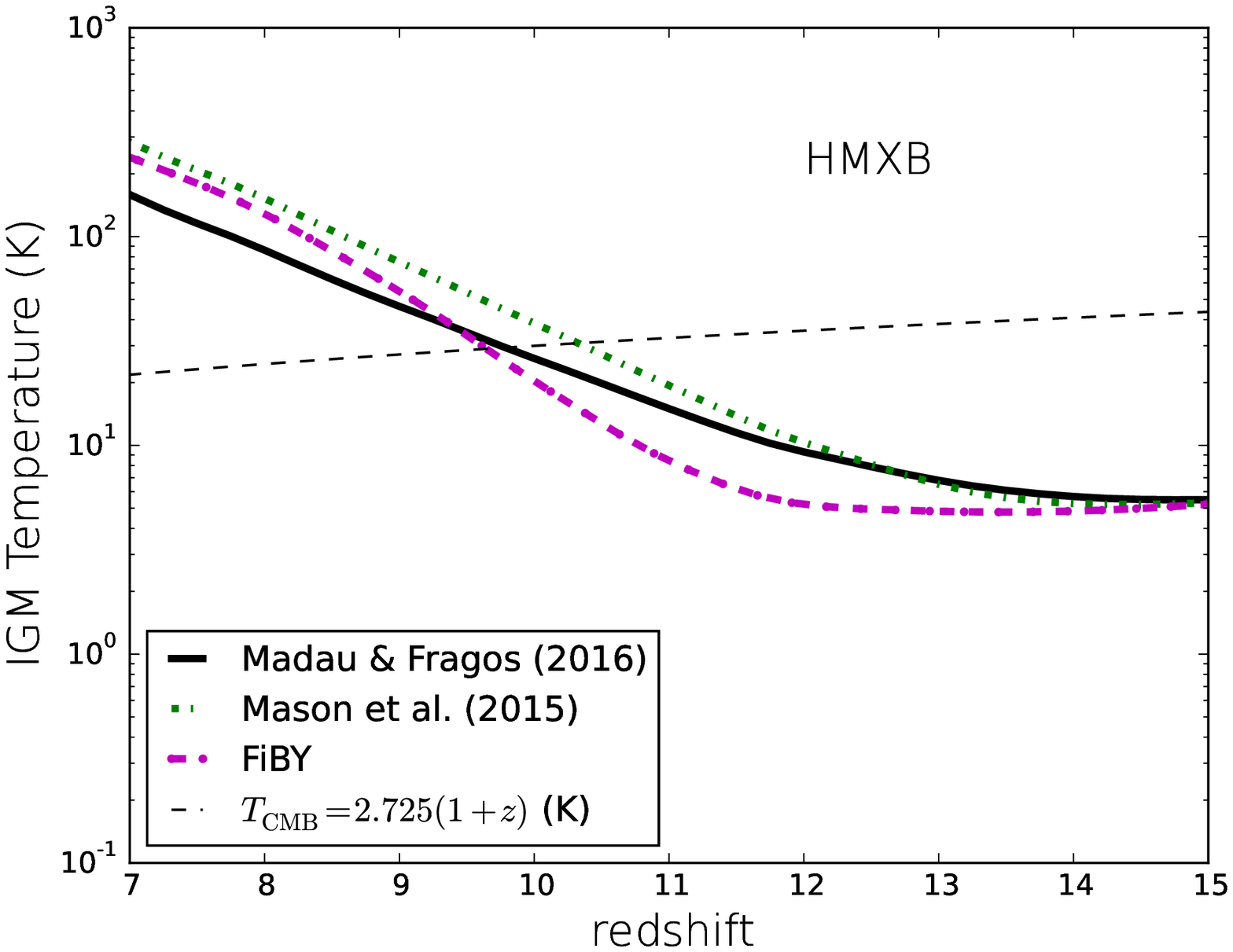}}
\caption{(left panel) Star formation rates based on \citet{2017ApJ...840...39M} (solid line), \citet{2015ApJ...813...21M} (green squares) and the FiBY simulation (blue points), normalized to a Salpeter stellar initial mass function. (right panel) The corresponding evolution of the IGM temperature, for an assumed X-ray heating efficiency of 36 percent (see text). 
}
\label{fig:HMXB}
\end{figure*}

The star formation rate histories from  \citet{2015ApJ...813...21M},
\citet{2017ApJ...840...39M} and the FiBY simulation, all normalized to
a common Salpeter stellar initial mass function for stars with masses
$0.1<M/M_\odot<100$, are shown in the left panel of
Fig.~\ref{fig:HMXB}. The corresponding temperatures to which the still
neutral IGM is heated by high mass X-ray binaries alone, computed by
scaling the heating rate from the results for the fiducial model of
\citet{2017ApJ...840...39M} according to the ratio of star formation
rates, are shown in the right panel. The varying star formation rate
histories produce only a small spread in the temperatures.


\bigskip
\section*{acknowledgments}

The authors thank Dr C. Mason for kindly providing look-up tables of the UV luminosity predictions as a function of halo mass based on the formalism of \citet{2015ApJ...813...21M}. The authors thank G. Wagstaff for computations of the X-ray emission in the Sedov-Taylor solution in preliminary work. JPP acknowledges support from the European Research Council under the European Community's Seventh Framework Programme (FP7/2007-2013) via the ERC Advanced Grant \lq\lq STARLIGHT:\ Formation of the First Stars\rq\rq (project number 339177).


\bibliographystyle{mn2e-eprint}
\bibliography{ms}

\label{lastpage}

\end{document}